\documentstyle[preprint,prd,aps,floats,epsfig]{revtex}

\def\real{I\negthinspace R}

\def\half{\textstyle{1\over2}}

\def\sech{{\rm sech}}
\newcommand{\e}{{\rm e}}
\newcommand{\mscr}[1]{{\mbox{\scriptsize #1}}}
\newcommand{\be}{\begin{equation}}
\newcommand{\ee}{\end{equation}}
\newcommand{\bea}{\begin{eqnarray}}
\newcommand{\eea}{\end{eqnarray}}
\newcommand{\bml}{\begin{mathletters}}
\newcommand{\eml}{\end{mathletters}}

\begin{document}
\preprint{DTP/99/5, gr-qc/9902081}

\draft
\tighten

\renewcommand{\topfraction}{0.8}

\title{Thick Domain Wall Universes}
\author{Filipe Bonjour,\footnote{E-mail address:
          \texttt{Filipe.Bonjour@durham.ac.uk}} Christos
        Charmousis,\footnote{E-mail address:
          \texttt{Christos.Charmousis@durham.ac.uk}}
        and Ruth Gregory\footnote{E-mail address:
          \texttt{R.A.W.Gregory@durham.ac.uk}}}
\address{Centre for Particle Theory, 
         Durham University, South Road, Durham, DH1 3LE, U.K.}
\date{\today}
\maketitle

\begin{abstract}
  
  We investigate the spacetime of a thick gravitating domain wall for a general
  potential $V(\Phi)$. Using general analytical arguments we show that all
  nontrivial solutions fall into two categories: those interpretable as an
  isolated domain wall with a cosmological event horizon, and those which are
  pure false vacuum de Sitter solutions.  Although this latter solution is
  always unstable to the field rolling coherently to its true vacuum, we show
  that there is an additional instability to wall formation if the scalar field
  does not couple too strongly to gravity. Using the $\lambda \Phi^4$ and
  sine-Gordon models as illustrative examples, we investigate the phase space
  of the gravitating domain wall in detail numerically, following the solutions
  from weak to strong gravity. We find excellent agreement with the analytic
  work. Then, we analyse the domain wall in the presence of a cosmological
  constant finding again the two kinds of solutions, wall and de Sitter,
  even in the presence of a negative cosmological constant.

\end{abstract}
\pacs{PACS numbers: 04.80.-b, 11.27.+d, 98.80.Cq \hfill DTP/99/5 \hfill
      gr-qc/9902081}


\section{Introduction}

The study of topological defects has wide applicability in many areas of
physics. In the cosmological arena, defects have been put forward as a possible
mechanism for structure formation~\cite{VS}, and while recent work on global
defects~\cite{ABR,PST} indicates that these were probably not responsible for
structure formation, the intriguing discovery of a non-gaussian signature in
the microwave background~\cite{FMG} leaves open the possibility that
defects were around at some point in the evolution of the universe.

Of all the topological defects, domain walls are the most deceptively simple to
study. They correspond to solitons in 1+1-dimensions, which are extended in two
spatial directions to form a wall structure.  Because they depend on only one
spatial coordinate, the distance from the wall, the solutions in the absence of
gravity can often be written in closed analytic form, and for certain
potentials the models are completely integrable and the defects have the full
interpretation of solitary travelling waves. In the presence of gravity
however, the situation changes, gravity destroys the integrability of the
theory, and also, except in a perturbative sense, the analytic nature of the
solutions.  Fortunately, since the walls correspond approximately to
hypersurfaces in spacetime, there is a well defined way of analysing their
gravity using Israel's thin wall formalism \cite{Israel}, in which the wall is
approximated by an infinitesimally thin hypersurface, and most of the
literature concerning domain wall gravity uses this method.

The domain wall is a rather interesting object gravitationally: unlike almost
all of the other topological defects (the exception being the global
string~\cite{R}), its metric is not static but time-dependent~\cite{Vil,IS},
having a de Sitter-like expansion in the plane of the wall. Observers
experience a repulsion from the domain wall, and there is an horizon at finite
proper distance from the defect's core.  This horizon can be interpreted as a
facet of the choice of coordinates, which usually use the flat space wall
solution as a starting point, and impose planar symmetry on the domain wall
spacetime. However, it is possible to use a different set of coordinates
\cite{IS,GWG}\ in which the wall has the appearance of a bubble which contracts
in from infinite radius to some minimum radius, then re-expands, undergoing
uniform acceleration from the origin. The `horizon' is then simply the
lightcone of the origin in these coordinates, and is somewhat similar to the
horizon of Rindler spacetime.

The crucial physical difference of a domain wall spacetime, as opposed to that
of the local cosmic string or monopole, is the presence of this cosmological
horizon, which introduces a second length scale into the system. Ordinarily, a
defect possesses one length scale, $w$, which is a measure of its thickness.
However, the distance to the event horizon of the domain wall gives another
length scale,
$z_{\rm h}$, which can be compared to $w$. Since these lengths are given in
terms of the coupling constants of the theory, taking a thin wall limit turns
out to be a very artificial construction in terms of these underlying
parameters, and the issue of the self-gravity of thick walls becomes more
pertinent. After the original work by Vilenkin, Ipser and
Sikivie~\cite{Vil,IS}, attempts focussed on trying to find a perturbative
expansion in the wall thickness~\cite{RD,larryw} both for the purpose of
discovering the motion of the wall, as well as verifying that the hypersurface
formalism was a good approximation to the true gravitational field. These
results backed up the thin wall approximation, the main difference being the
presence of a sub-dominant tension transverse to the wall.

With the suggestion of Hill, Schramm and Fry~\cite{HSF} of a late time phase
transition with thick domain walls, there was some effort at finding exact
thick solutions~\cite{Goetz,Mukh}, see also \cite{Tom,TI}.  However, such
defects were supposed to be thick by virtue of the low temperature of the phase
transition responsible for defect formation. The suggestion that Planck scale
topological defects could be responsible for inflation \cite{Linde,Vil2} then
reopened the issue of thick domain walls. `Thick' then means thick compared to
their natural de Sitter horizon, and therefore this appears to be a very strong
gravity situation. The issue of whether a domain wall can survive in a
Friedmann-Robertson-Walker (FRW) universe with horizon size comparable to wall
thickness has been analysed in some detail for the case of no gravitational
back reaction of the wall on the FRW background~\cite{BV1,BBGH}, and in the
case of Euclidean instantons on a de Sitter background including self-gravity
\cite{BV2,BGV}; 
however, to our knowledge, a systematic analysis of the strong self-gravity 
of thick domain walls with an arbitrary field theory potential has 
not been carried out.

In this paper we perform a detailed analysis of the strong gravity of thick
domain walls. With a combination of analytical and numerical results, we map
out the parameter space in which a wall-like solution exists. Although we focus
on two main examples of the sine-Gordon and $\lambda\Phi^4$ wall, we also
illustrate to what extent, and for what potentials, the results will also hold
in general.  We begin in the next section by setting out the general formalism
for a thick domain wall, deriving the metric and Einstein equations, as well as
defining what we mean by a `wall' solution. We also show how to generalise a
coordinate transformation which takes a planar domain wall spacetime with an
horizon into a flat spacetime with an accelerating wall to the case of walls
with finite thickness. The solution outside the wall horizon is shown to be
related to an FRW cosmology with a slowly rolling scalar field.

In the following section we derive analytic results for the thick wall. It
turns out that there are two possibilities for the scalar field from which the
`wall' is constituted: it can either be a planar wall or a de Sitter solution.
Since the de Sitter solution is always possible, but not necessarily stable, we
derive analytic bounds on the gravitational coupling strength of the wall for
when the de Sitter solution is the only possible solution, and when it becomes
unstable to decay into a wall-like solution. Note that the de Sitter solution
\emph{always} contains an instability to the field rolling coherently down the
potential well~--- we will not be interested in these instabilities, only in
those which would lead to the formation of a domain wall. We derive these
bounds for both our chosen field theory models, as well as for a general
potential. We then provide, in the case of weak gravitational coupling, some
perturbative solutions for the wall and its gravitational fields. These can be
readily compared to existing results in the literature. We then present
numerical results backing up and extending the analytic work. In the
penultimate section we consider the domain wall in the presence of a 
cosmological constant, and conclude in the final section.

\section{Plane-symmetric spacetimes}\label{sec2}

We start by setting up the general framework for a domain wall coupled to
gravity. We initially consider a general matter Lagrangian
\be
{\cal L}_\mscr{M} = ( \nabla_a \Phi )^2 - U \left ( \Phi \right ),
\label{wallag}
\ee
where $\Phi$ is a real Higgs field and the symmetry breaking potential
$U(\Phi)$, has a discrete set of degenerate minima. We assume that the spacing
of these minima is proportional to the (dimensionful) parameter $\eta$, which
sets the symmetry breaking scale, and that $U(\Phi)$ is characterised by a
scale $V_\mscr{F} = U(\Phi_\mscr{F})$, where $\Phi_\mscr{F}$ is a local false
vacuum situated between successive minima ($\Phi_\mscr{F} = 0$ is a
conventional choice). For example, in the usual `kink' model, $U(\Phi) =
\lambda \left ( \Phi^2 - \eta^2 \right )^2$, and we see $\eta$ directly, with
$V_\mscr{F} = \lambda \eta^4$. For convenience, we scale out dimensionful
parameters via
\be \label{epsilon}
  X = \Phi/\eta, \qquad \epsilon = 8\pi G\eta^2.
\ee
The dimensionless parameter $\epsilon$, which we call gravitational strength
parameter, characterises the gravitational interaction of the Higgs field.
Then, defining $V(X) = U(\eta X)/V_\mscr{F}$,\footnote{Note that $V(X_{\rm
F})=1$ by definition.}
\be
8\pi G{\cal L}_\mscr{M} = {\epsilon \over w^2} \left [
w^2 (\nabla_a X)^2 - V(X) \right ],
\label{rescaledlag}
\ee
where $w = \sqrt{\epsilon \over 8\pi G V_\mscr{F}}$ represents the inverse mass
of the scalar after symmetry breaking, and of course will also characterise the
width of the wall defect within the theory.  The equations of motion following
from (\ref{rescaledlag}) are simply
\be
\Box X + {1 \over 2w^2} {\partial V \over \partial X} = 0.
\label{walleqs}
\ee
Without loss of generality we can set $w=1$ (which amounts to choosing `wall'
units rather than Planck units) and, looking for a static solution in flat
space, we see that (\ref{walleqs}) can be integrated directly to give
\be
X'^2 = V(X),
\label{flatnlde}
\ee
which has an implicit solution
\be
\int^{X}_{X_\mscr{F}} {dX\over\sqrt{V(X)}} = z - z_0,
\label{flatx}
\ee
where $X_\mscr{F} = X(z_0)$ is the false vacuum. For example, in the
$\lambda\Phi^4$ model above, $z-z_0 = \int_0^X dX/(1-X^2) = \tanh^{-1} X$, and
we get the usual kink solution centered on $z_0$: $X = \tanh(z-z_0)$.  Another
model that we will be exploring in detail is the sine-Gordon model, with
$V(\Phi) = V_\mscr{F} [1 + \cos (\Phi/\eta) ]/2$, in which case $z - z_0 =
2\ln\tan (\pi/4 + X/4)$.

We now look for a plane-symmetric gravitating domain wall solution, since this
represents the most obvious intuitive generalization of the flat space domain
wall. We will consider coordinate transformations of this solution at the end
of the section.  The metric therefore will have planar symmetry (i.e. Killing
vectors $\partial_x, \partial_y, x\partial_y - y\partial_x$), and in addition
will display reflection symmetry around a surface, $z=0$ say, which represents
the location of the wall (defined by $X = X_{\rm F}$). If we choose $z$ to be
the proper distance from the wall, then the metric may be written as
\be
\label{met1}
ds^2 = A^2(z) dt^2 - B^2(z,t) \left( dx^2+dy^2 \right) - dz^2,
\ee
with the associated Einstein equations derived from the
Lagrangian~(\ref{wallag}) coupled to gravity through the usual Hilbert term,
${\cal L}_\mscr{G} = - R/ 2\epsilon$, being
\be 
  R_{ab} = 2\epsilon X_{,a} X_{,b} - \epsilon \, V(X) \, g_{ab}.
  \label{eeqs}
\ee
Before writing these equations explicitly, we will first examine what we mean
by a domain wall solution, since this will require some rather specific
boundary conditions at $z=0$, and for large $z$. We will define a \emph{wall
solution} to be a function $X(z)$ of the proper distance from the wall which
at $z=0$ is at a local maximum of $V(X)$, $X_\mscr{F}$, and which falls towards
distinct minima on either side of the wall. Assuming that $V(X)$ is locally
symmetric around the maximum, then $(X(z) - X_{\rm F})$ will be an odd function
of $z$. This restriction embodies the idea of a defect, in which the field
falls to distinct vacua on either side of its core, and settles to a
topological, rather than radiative, configuration. It is possible that the only
nonsingular solution satisfying these criteria is $X \equiv X_\mscr{F}$, in
which case the spacetime will be de Sitter; indeed, this is always a possible
solution to the equations of motion satisfying the above criteria, though it
will not necessarily be stable.  Clearly, there is always an instability
corresponding to a coherent roll of the field towards one of its vacuum values,
however, since we are interested in spacetimes with the interpretation of a
domain wall, we will ignore this mode, and say that de Sitter spacetime is
`stable' if there is no odd perturbation of $(X - X_{\rm F})$ which is growing
in time. As we will see, the scalar field does not fall all the way to its
minimum value within the range of validity of the coordinates in (\ref{met1}),
which is why we do not place any specific boundary conditions on $X$ for large
$z$, however, it will turn out that we require $X'(z_{h}) = 0$ for a
nonsingular solution.  Finally, we can choose $t$ to set $A(0) = 1$, and
reflection symmetry requires $A'(0)=0$.

Turning to the Einstein equations (\ref{eeqs}), we see that
\be
R_{zt} = {{\dot B} A'\over BA} - {{\dot B}' \over B} = 0
\ee
which implies $B = b(t) A(z)$. Then the relation
\be
R^t_t - R^x_x = A^2 \left ( {{\dot b}^2 \over b^2} - {{\ddot b}\over b}
\right ) = 0
\ee
yields
\be
b(t) = \e^{kt},
\ee
so that the equations of motion for the gravitating wall finally reduce simply
to
\bml\bea
X'' + 3 {A'\over A} X' &=& {1\over2} {\partial V\over \partial X} 
\label{gweqb} \\
{A''\over A} &=& -{\epsilon\over3} \left [ 2 X^{\prime 2} + V(X) \right ] 
\label{gweqa}\\
\left ( {A'\over A} \right )^2 &=& {k^2\over A^2} + {\epsilon\over3}
\left [ X^{\prime 2} - V(X) \right ]. \label{gweqc} 
\eea\label{gweqs}\eml
Note that since $A''\leq 0$, once $A'$ becomes negative it will always be
bounded away from zero, therefore there is some finite $z_{\rm h}$ for which
$A(z_{\rm h})=0$, and we have either a physical or coordinate singularity.
Thus we see immediately that there is no nonsingular solution with $k^{2}=0$.
It is also easy to see that the false vacuum de Sitter solution is given by $X
= X_\mscr{F}$, $A=\cos kz$, $k^2 = \epsilon/3$.\footnote{This somewhat less
familiar form is merely one of the many coordinate transformations of de
Sitter, and can be reduced to the more familiar form $ds^2 = d\tau^2 -
\e^{2k\tau} d{\bf x}^2$ via the transformation $\e^{k\tau} = \e^{kt}\cos kz$,
$\zeta = \tan(kz) \e^{-kt} /k$.}

Determining therefore whether or not a wall solution exists reduces to
investigating the system of equations (\ref{gweqs}). Clearly, for small
$\epsilon$ we might expect a wall solution to the above equations to exist, and
to be given by a perturbative expansion around flat space. For large
$\epsilon$, since $A' = O(\epsilon)$, the $A'X'/A$ term in (\ref{gweqb}) will
drive the solution to a singularity for nonzero $(X-X_\mscr{F})$, hence 
for large $\epsilon$ we expect only the de Sitter solution. For intermediate
values of $\epsilon$, except for some very special cases, an
analytic solution does not exist and we have to numerically integrate the
equations.

Before proceeding with the details of this analysis, we conclude this section
by commenting on a coordinate transformation of the plane symmetric metric
(\ref{met1}), which transforms the defect into an accelerating bubble wall.
Defining
\bml\label{starred}\bea
  x^* &=& A(z) \e^{kt} x \\
  y^* &=& A(z) \e^{kt} y \\
  t^*-z^* &=& -{1\over k} A(z) \e^{kt} \\
  t^*+z^* &=& \phantom{-}{1\over k} A(z) \e^{-kt} - k(x^2+y^2) A(z) \e^{kt}
\eea\eml
gives an alternate form of the line element:
\be
  \label{starmet}
  ds^2 = dt^{*2} - dx^{*2} - dy^{*2} - dz^{*2} + \left ( {A^{\prime2}\over k^2}
  -1 \right ) dz^2,
\ee
where $z$ is given implicitly by
\be
  k^2 (t^{*2} - {\bf x}^{*2}) = - A^2(z).
\ee
The wall (i.e.\ the zero of $X-X_{\rm F}$) is located at ${\bf x}^{*2} - t^{*2}
= 1/k^2$, and as we will see, for small values of the gravitational coupling
$\epsilon$, $A'$ rapidly approaches $k = O(\epsilon)$ outside the core of the
wall. Therefore, we see that the spacetime in these coordinates is
approximately flat, with the wall being located at a spacelike hyperboloid at a
distance $1/k$ from the origin. This corresponds to a wall undergoing uniform
acceleration, contracting in from infinity, reaching a minimum radius, and
re-expanding outwards. The horizon ($A(z_{\rm h})=0$ in the old coordinates) is
now the lightcone centered on the origin, and the region exterior to the
horizon is the causal future and past of the origin.  Thus the coordinate
transformation (\ref{starred}) generalises the thin wall transformation given
in \cite{IS}, and discussed in detail in section three of \cite{GWG}.

For larger values of $\epsilon$ the spacetime will not be flat, and $X$
will not be at its true vacuum value near the horizon. In particular, setting
$t^* = \tau\cosh\psi$, $x^* = \tau \sinh \psi \sin \theta \cos\phi$, etc., then
gives a rather familiar form for the metric exterior to the horizon, i.e.\ 
inside the lightcone of the origin:
\be
  ds^2 = C^2(\tau) d\tau^2 - \tau^2 \left [ d\psi^2 + \sinh^2\psi (
  d\theta^2 + \sin^2\theta d\phi^2 )\right ],
\ee
the metric of an open FRW universe. Thus if the scalar field is not at its
vacuum value at the horizon, its evolution outside the horizon is given by the
rolling of a scalar towards its minimum in an open FRW model, a well studied
problem!

\section{self-gravitating domain walls}\label{sgwall}

We now want to find solutions to~(\ref{gweqs}) representing an isolated domain
wall for various values of the parameter $\epsilon$. We start by considering a
general potential, $V(X)$, finding an analytic perturbative solution for small
$\epsilon$, and showing that if $\epsilon$ is sufficiently large there is no
wall solution. We also demonstrate an instability of the false vacuum de Sitter
solution to wall formation for $\epsilon \leq \epsilon_{\rm max}$, where
$\epsilon_{\rm max}$ depends on the second derivative of the potential at the
origin. We then explore the particular cases of the $\lambda \Phi^4$ kink and
the sine-Gordon model in some detail, first making reference to the analytic
work, then describing the solutions for intermediate values of $\epsilon$ with
the help of numerical work. Both analytic and numerical work show the presence
of a phase transition in the behaviour of the solutions between wall existence
and nonexistence.

It is clear that when the gravitational strength parameter is set to zero, one
gets the flat space solution $A=1$, with $X$ being given by (\ref{flatx}). Let
us now consider small values of $\epsilon$, typically $\epsilon\ll 1$. Then we
can expand the fields $X$ and $A$ in powers of $\epsilon$
\bml\bea
  X &=& X_0 + \epsilon X_1 + O\left(\epsilon^2\right)\\
  A &=& A_0 + \epsilon A_1 + O\left(\epsilon^2\right),
\eea\eml
where $X_n$, $A_n$ are now independent of $\epsilon$ and $A_0, X_0$ are the
flat space solutions.

The field equations (\ref{gweqs}) to lowest order in $\epsilon$ give
\bml\bea
  A_1'' &=& -{1\over3} \left [ 2 X_0^{\prime2} + V(X_0) \right ] 
  =-X_0^{\prime2} \label{onea}\\
  X_1'' &=& - 3 A_1' X_0' + {1\over 2} X_1 {\partial ^2 V \over \partial X^2 }
  \bigg | _{X_0(z)}. \label{oneb}
\eea\eml
The boundary conditions of $A$ and $X$ give
\be
  A_1(0) = A_1'(0) = 0, \qquad X_1(0) =0, \qquad X_{1} \to 0 \mbox{ for large
  }z; \label{fobcs}
\ee
(\ref{onea}) and~(\ref{oneb}) can be integrated to give
\be
  A_1 =  - \int \!\!\!\! \int V(X_{0}) dz =  - 
  \int {dX\over \sqrt{V}}\int\sqrt{V}dX
  \label{aone}
\ee
and
\be
  X_1 = - {3\over2} X_0' \int {dz\over X_0^{\prime2}} \left ( A_1'^{2}- 
  {k^{2}\over \epsilon^{2}} \right ) dz,
  \label{xone}
\ee
which can also be expressed as an implicit integral in terms of $V(X)$ and $X$.
Finally, noting that
\be
  X_0'X_1' - X_1 X_0'' = {3\over2} A_1^{\prime2} + X_0'(0) X_1'(0),
\ee
(\ref{gweqc}) implies
\be
  k^2 = \epsilon^2 \left [ A_1^{\prime2} + {2\over3} (X_1X_0'' - X_0'X_1')
  \right ] = - {2\epsilon^2\over3} X_0'(0) X_1'(0).
  \label{kone}
\ee

Before exploring these solutions for specific models for $V(X)$, we can make
some general statements about the self-gravitating wall. First of all, since
$A_1''$ is strictly negative, $A$ will asymptote $1 - kz$, where $k$ is
O($\epsilon$) from the above relation. This means that $A_1$ will cease to be
small at a distance of order $\epsilon^{-1}$ from the wall, and our expansion
procedure strictly speaking breaks down. However, since $A_1'$ is not growing,
it is clear that continuing the expansion to higher orders in $\epsilon$ will
merely produce minor corrections to $A$, and will not alter the qualitative
behaviour, namely, that $A(z)$ has a zero at a distance of order
$\epsilon^{-1}$ from the wall (see figure~\ref{fig:comp}). Since $g_{tt}=A^2$
this is simply the event horizon which is familiar from the thin wall
approximation.

As $\epsilon$ increases, the effect of the scalar field's energy-momentum on
the geometry increases, and we expect the horizon to move closer to the wall,
roughly until $\epsilon = {\rm O}(1)$. For large $\epsilon$, the expected
horizon would be well inside the core of the wall and two possibilities now
emerge.  First of all, the scalar field could simply ignore the geometry, and
fall away minutely from its false vacuum, causing an horizon at some small
value of $z$.  Since the spatial gradient of such a solution would be
relatively small, the energy-momentum would be vacuum dominated, and we would
expect the spacetime to be very close to the de Sitter solution. Outside the
horizon, the scalar field would roll to its vacuum value as described in the
last section. 
The alternative
is that there is a phase transition in the behaviour of $X$, that is, that $X$
either has some nontrivial odd form, approaching reasonably close to its vacuum
value at the horizon, or $X \equiv X_\mscr{F}$. In other words, $X$ must either
roll significantly away from $X_\mscr{F}$, or not at all. There are two reasons
why we expect the latter scenario to hold. The first is that Basu and Vilenkin
\cite{BV2} observed just such a phase transition in studying the problem of
wall instantons.
The other reason for suspecting a phase transition lies in the behaviour of
field theory solutions on compact surfaces. For example, two of the authors
have studied a cosmic string interacting with the event horizon of an extremal
black hole~\cite{BEG}.  There, there are nontrivial solutions with the string
piercing the horizon while the string fields can fall reasonably close to their
vacuum values around the horizon; however, there is a transition when the
string becomes sufficiently thick relative to the black hole: the event horizon
can no longer support a nontrivial solution and the flux of the string is
expelled~--- the only horizon solution is the trivial one.

First of all, let us show that for small $\epsilon$ the false vacuum de Sitter
solution is unstable. Recall that the de Sitter solution is
\be
\label{sitter}
ds^2 = \cos^2(kz) dt^2 - \e^{2kt} \cos^2(kz) (dx^2+dy^2) - dz^2
\ee
with $X \equiv X_\mscr{F}$ and $k^2 = \epsilon/3$. This solution will be
`stable' (i.e., stable to wall formation) if there is no perturbation of $X -
X_{\rm F}$ which is an odd function of $z$ and growing in time. Setting $X =
X_\mscr{F} + \xi$, and noting that corrections to the geometry are O($\xi^2$),
we see that an instability must satisfy the time-dependent linearized
perturbation equation
\be
\label{stab2}
\xi'' - 3k\tan(kz) \xi' - \sec^2(kz) [ \ddot{\xi} + 2k \dot{\xi} ] -
\frac12 \xi {\partial^2 V\over \partial X^2} \bigg | _{X_\mscr{F}} = 0
\ee
(where a dot indicates differentiation with respect to $t$). This equation does
indeed have unstable solutions for $k^2 = \epsilon/3 <|V''(X_{\rm F})|/8$, the
dominant instability being given by
\be
\xi = \e^{k\nu t} \sin kz (\cos kz) ^\nu
\ee
with
\be
\nu = -{5\over2} + {1\over2} \sqrt{ 9 - 2V''(X_\mscr{F})/k^2}.
\label{nudef}
\ee
Thus for $\epsilon$ smaller than
\be\label{epsmax}
  \epsilon_{\rm max} = \frac38 \; \left|V''(X_\mscr{F})\right|,
\ee
the de Sitter solution is unstable to wall formation.

Now let us examine whether a wall solution can exist for large $\epsilon$.  For
a nontrivial wall solution, we require $X'(0)>0$, and for a nonsingular
solution, we require $X'(z_{\rm h})=0$.  Now taking the derivative of the field
equation (\ref{gweqb}) we get
\be
  X'''= -3{A' \over A}X''+X'\left[-3{A'' \over A}+3\left({A' \over A}\right)^2
  + \frac12 {\partial^2V\over\partial X^2} \right]
  = -3{A' \over A}X''+X' F(z) 
  \label{nowall1}
\ee
where $F(z)$ may be rewritten using~(\ref{gweqs}) as
\be
F(z) = \left[ 3\epsilon X^{\prime2} + 3{k^2\over A^2}
+ \frac12 {\partial^2V\over\partial X^2} \right].
\ee
Now, at $z=0$ (\ref{gweqc}) gives $3k^2 = \epsilon[1-X'(0)^2]$, hence
\be
X'''(0) = X'(0) \left [ 3\epsilon - 6 k^2 + \frac12 V''(X_\mscr{F}) \right]
> X'(0) \left [ \epsilon + \frac12 V''(X_\mscr{F}) \right]
\ee
Therefore, if $\epsilon > |V''(X_\mscr{F})|/2$, $X'''(0)>0$, and $X'$ is
increasing away from $z=0$. Moreover, if $|V''(X)|$ is maximized at
$X_\mscr{F}$, then $F(z)$ is strictly increasing away from $z=0$ and $X'$ can
never be zero at the horizon. Thus with a minimal assumption on the nature of
the potential, we have shown that there is no wall solution possible if
\be
\epsilon > \frac12 \, |V''(X_\mscr{F})|.
\label{esup}
\ee
This argument does not make any assumptions as to the behaviour of the
geometry, it simply relies on general properties of the potential.

To reiterate, we have shown that for $\epsilon \leq \epsilon_{\rm max}$ there
are two solutions: a wall spacetime and a de Sitter solution, the latter
being unstable to wall formation.  For $\epsilon > \epsilon_{\rm max}$, 
the de Sitter solution is `stable', and for $\epsilon > 4\epsilon_{\rm max}/3$, 
the domain wall solution can analytically be shown not to exist. To
determine whether de Sitter is the only solution for $\epsilon_{\rm max} <
\epsilon < 4\epsilon_{\rm max}/3$, the problem must be examined numerically.
We now do this, and obtain explicitly the perturbative solution for the
$\lambda\Phi^4$ and sine-Gordon potentials.

\subsection{The $\lambda\Phi^4$ kink}

\begin{figure}[hbtp]
  \centerline{\epsfig{file=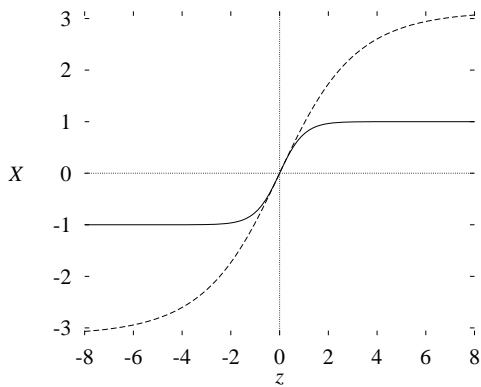,width=6.8cm}}
  \vspace*{0mm}
  \caption{The flat spacetime solution. The solid line is $X = \tanh(z)$, the
           solution for the Goldstone model, and the broken line is $X = 4
           \arctan \left( \e^{\frac z2} \right) - \pi$, the solution for the
           sine-Gordon case of section~\ref{sec:sG}.}
  \label{fig:flatwall}
\end{figure}

In this case the potential $V(X)$ is given by $ V(X)=(X^2-1)^2 $, and in flat
spacetime (O$\left(\epsilon^0\right)$) we have the usual flat domain wall
solution plotted in figure~\ref{fig:flatwall},
\be
  X_0 = \tanh(z).
\ee

\begin{figure}[htbp]
  \centerline{\epsfig{file=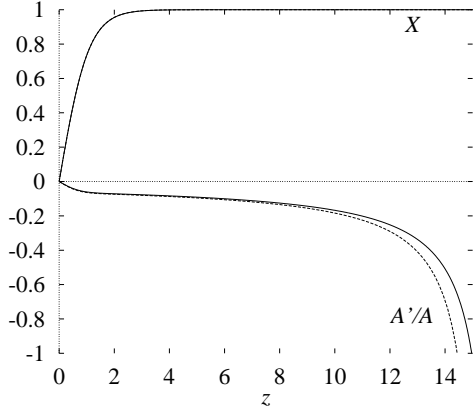,width=6.8cm}}
  \caption{Comparison between the solution obtained numerically (solid line)
           for $\epsilon = 0.1$ and the series to order O($\epsilon$)
           [equation~(\protect\ref{eq:wallseries1})]. (The two solutions for
           $X$ appear identical at this scale.)}
  \label{fig:comp}
\end{figure}

Integrating (\ref{aone}), (\ref{xone}) and calculating (\ref{kone}) one gets,
\bml\bea
  X_1 &=& -{1\over2} \sech^2 z \left [ z + {1\over 3} \tanh z \right ] \\
  A_1 &=& -{2\over 3} \log\cosh z - {1\over6}\tanh^2 z \\
  k   &=& \phantom{-}{2 \over 3}\epsilon+{\rm O}\left(\epsilon^2\right),
\eea\eml
which represent the first order gravitational corrections to $X$ and $A$. Note
also that $X_1$ does indeed satisfy the boundary conditions (\ref{fobcs}). The
distance to the event horizon is given by $z_{\rm h} \simeq 3 / 2 \epsilon$.
Putting together our results we get to order O$(\epsilon)$
\begin{mathletters}
\label{eq:wallseries1}
\begin{eqnarray}
  X &=& \tanh z - \frac \epsilon 2 \sech^2 z \left[ z + \frac13 \tanh z
        \right] \\
  A &=& 1 - \frac \epsilon 3 \left[ 2\ln \cosh z + \frac 12 \tanh^2 z
        \right].
\end{eqnarray}
\end{mathletters}
This solution is compared on figure~\ref{fig:comp} with the one found
numerically, for $\epsilon = 0.1$.

\begin{figure}[htbp]
  \begin{center}
    \epsfig{file=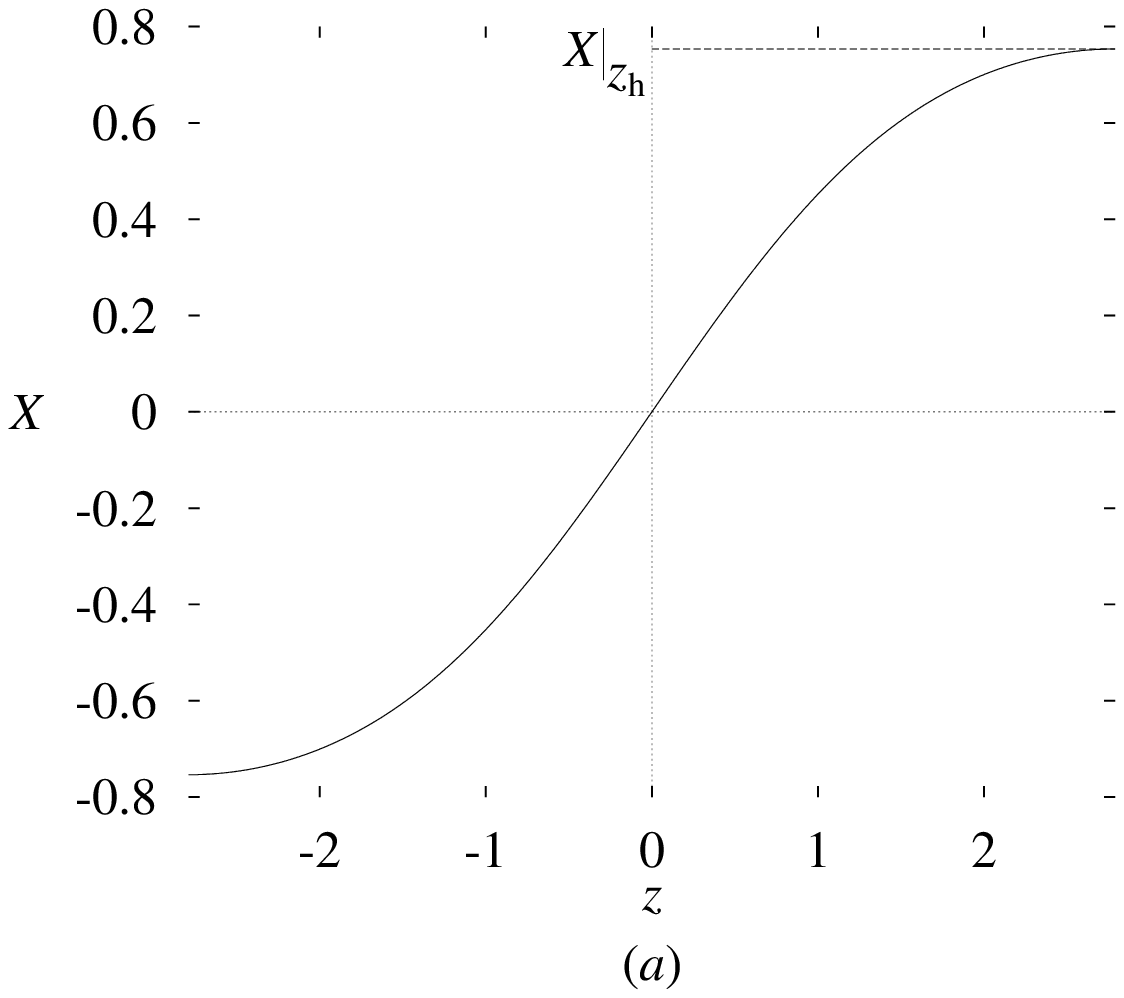,width=6.8cm} \qquad
    \epsfig{file=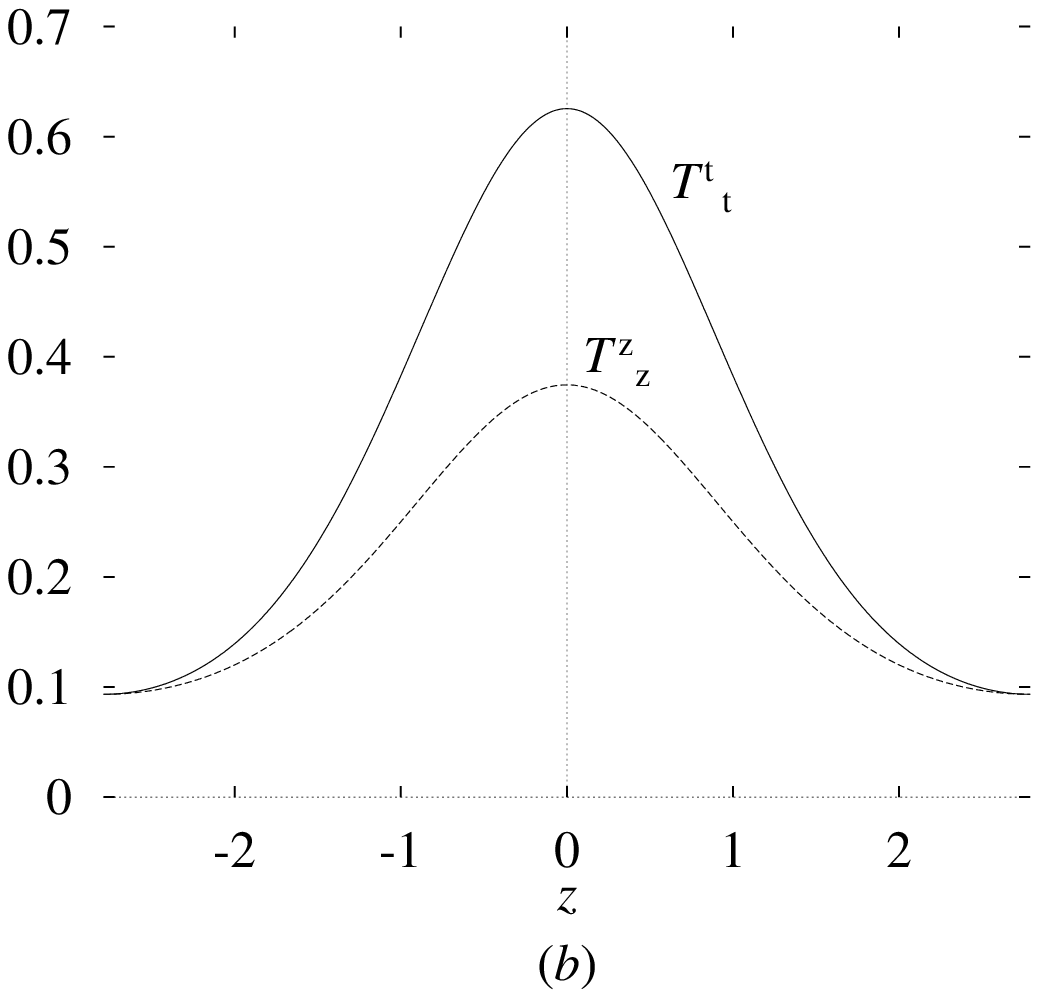,width=6.8cm} \\
    \epsfig{file=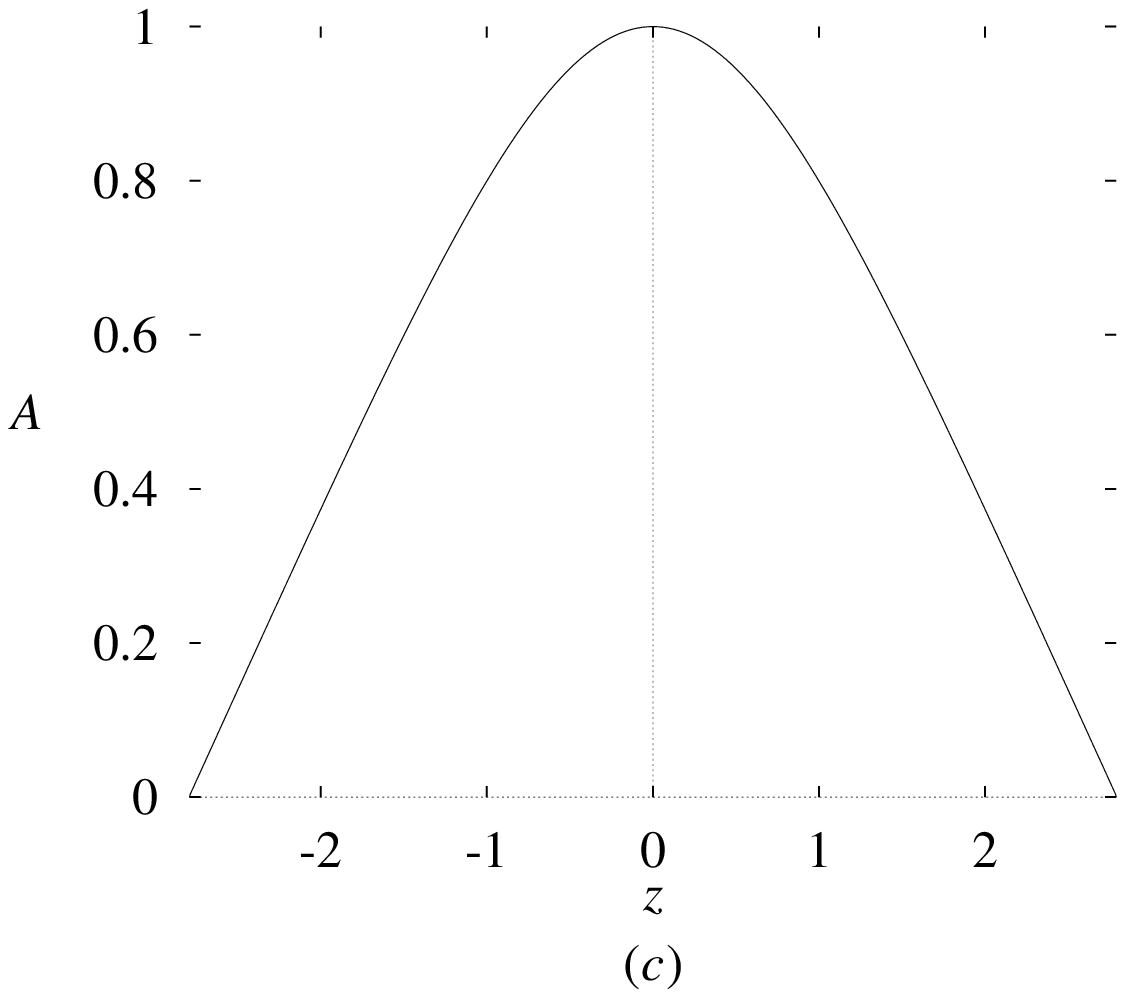,width=6.8cm}   \qquad
    \epsfig{file=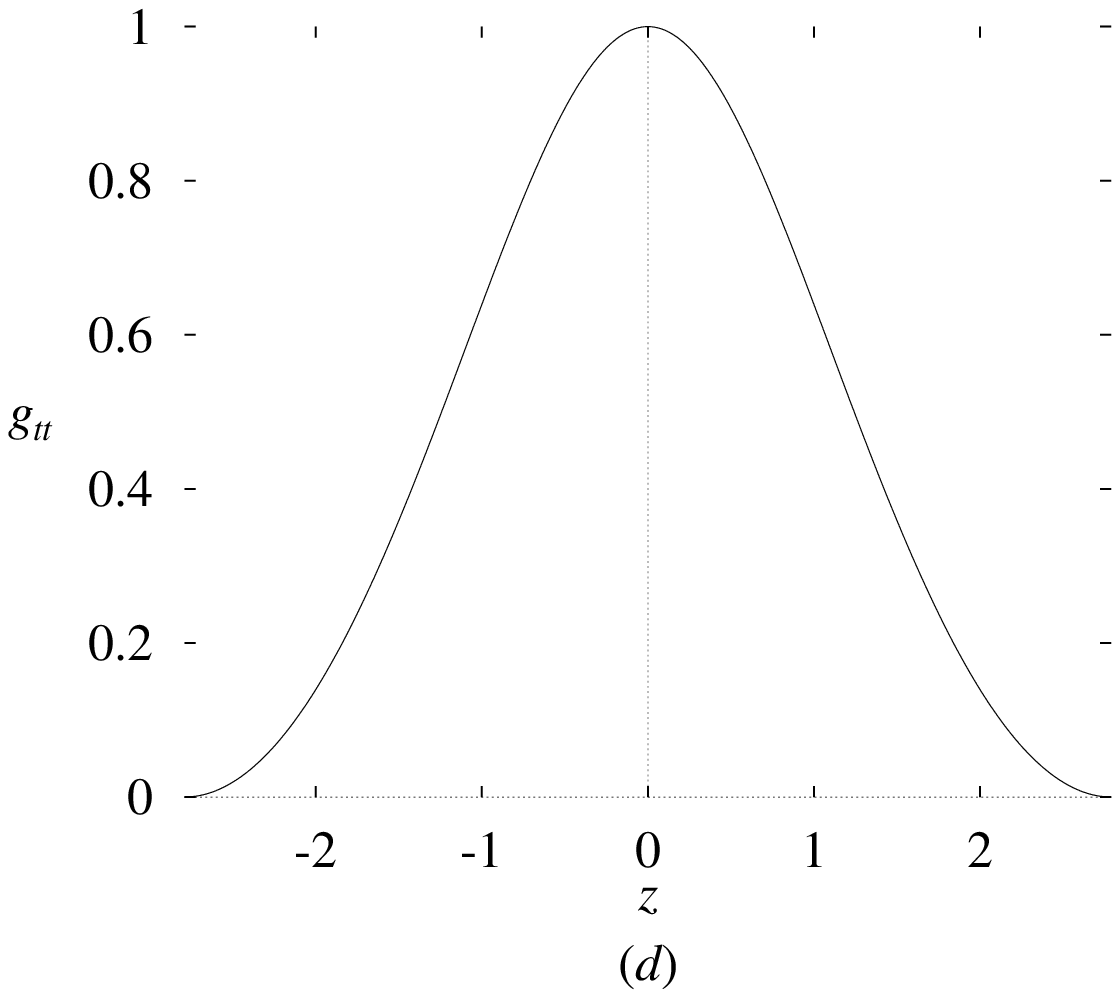,width=6.8cm}
  \end{center}
  \caption{Numerical solution of the equations~(\protect\ref{gweqs}) for the
           $\lambda \Phi^4$ model. This solution was obtained for $\epsilon =
           0.9$ (in which case the horizon was situated at a proper distance
           $z_\mscr{h} = 2.789$). The figure shows~({\it a\/}) the Higgs field,
           ({\it b\/}) the energy momentum tensor $T^x{}_x = T^y{}_y = T^t{}_t$
           and $T^z{}_z$, ({\it c\/}) the function $A(z)$ and~({\it d\/}) the
           metric component $g_{tt}(z) = A^2(z)$.}
  \label{fig:soln}
\end{figure}

For larger values of $\epsilon$, we must resort to numerical methods to find
solutions of~(\ref{gweqs}). Here, we have used the routine \textsc{solvde}
from~\cite{NumRec}. The wall solutions that we obtain are qualitatively the
same as the one shown on figure~\ref{fig:soln}. Note that as mentioned
previously, $X$ does not go to its asymptotic value at $z_{\rm h}$ (and
consequently that the energy density does not tend to zero at the horizon). In
fact, we do not solve~(\ref{gweqs}) as written in section~\ref{sec2};
instead, we rewrite equation~(\ref{gweqa}) as
\be
  \left( \frac{A'}A \right)' + \left( \frac{A'}A \right)^2 + \frac \epsilon 3
    \left[ 2X'{}^2 + V(X) \right] = 0.
\ee
The system~(\ref{gweqb},~\ref{gweqa}) can now be written as three coupled first
order ordinary differential equations (ODE's),
\bml\bea
  X' &=& Y \\
  Y' &=& - 3 Y Z + \frac12 \frac{\partial V}{\partial X} \\
  Z' &=& - \frac \epsilon 3 \left[ 2Y{}^2 + V(X) \right] - Z^2,
\eea\eml
where $Z = A'/A$. These equations were solved for the boundary conditions
$X(0) = Z(0) = Y(z_\mscr{h}) = 0$.

\begin{figure}[htbp]
  \begin{center}
    \epsfig{file=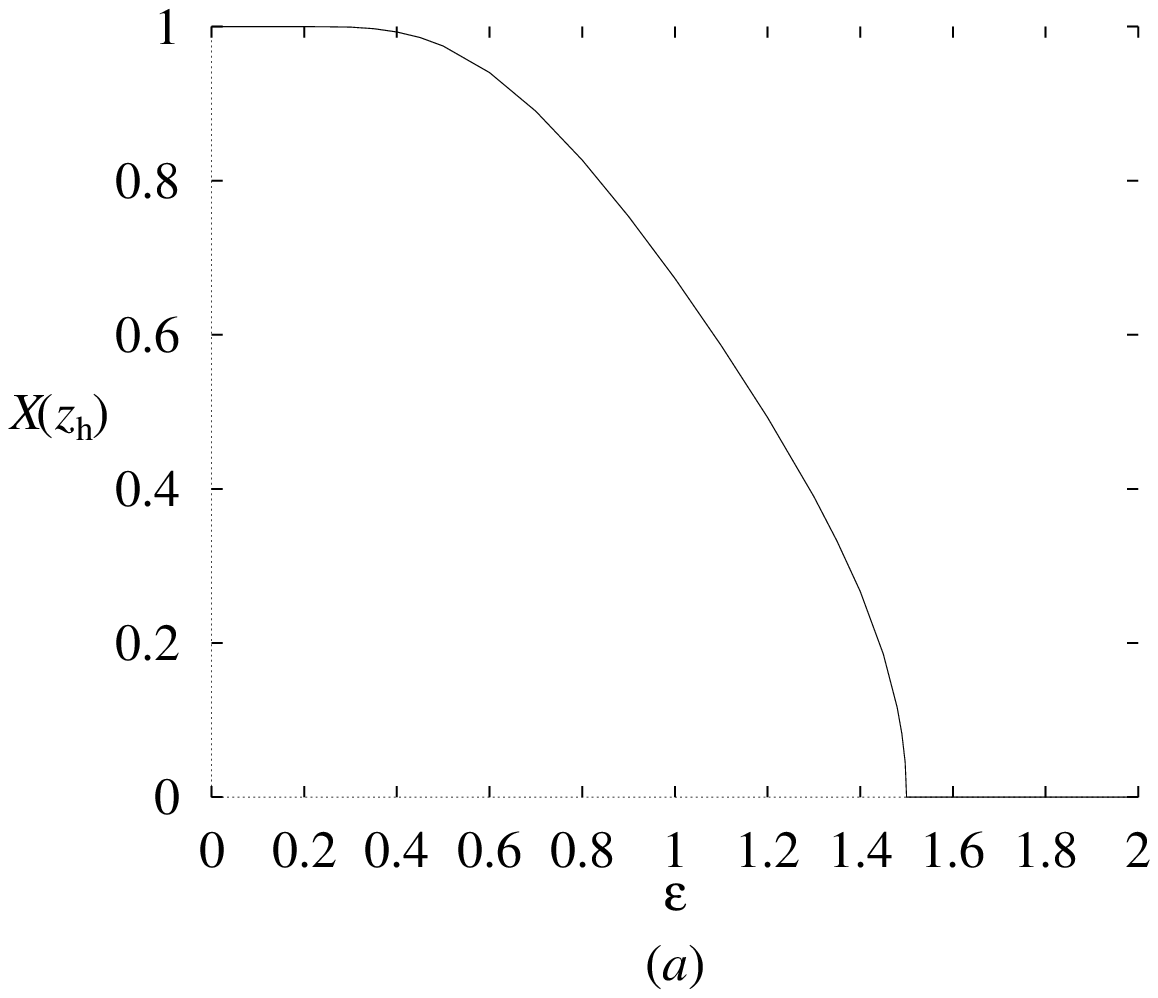,width=6.8cm}
    \epsfig{file=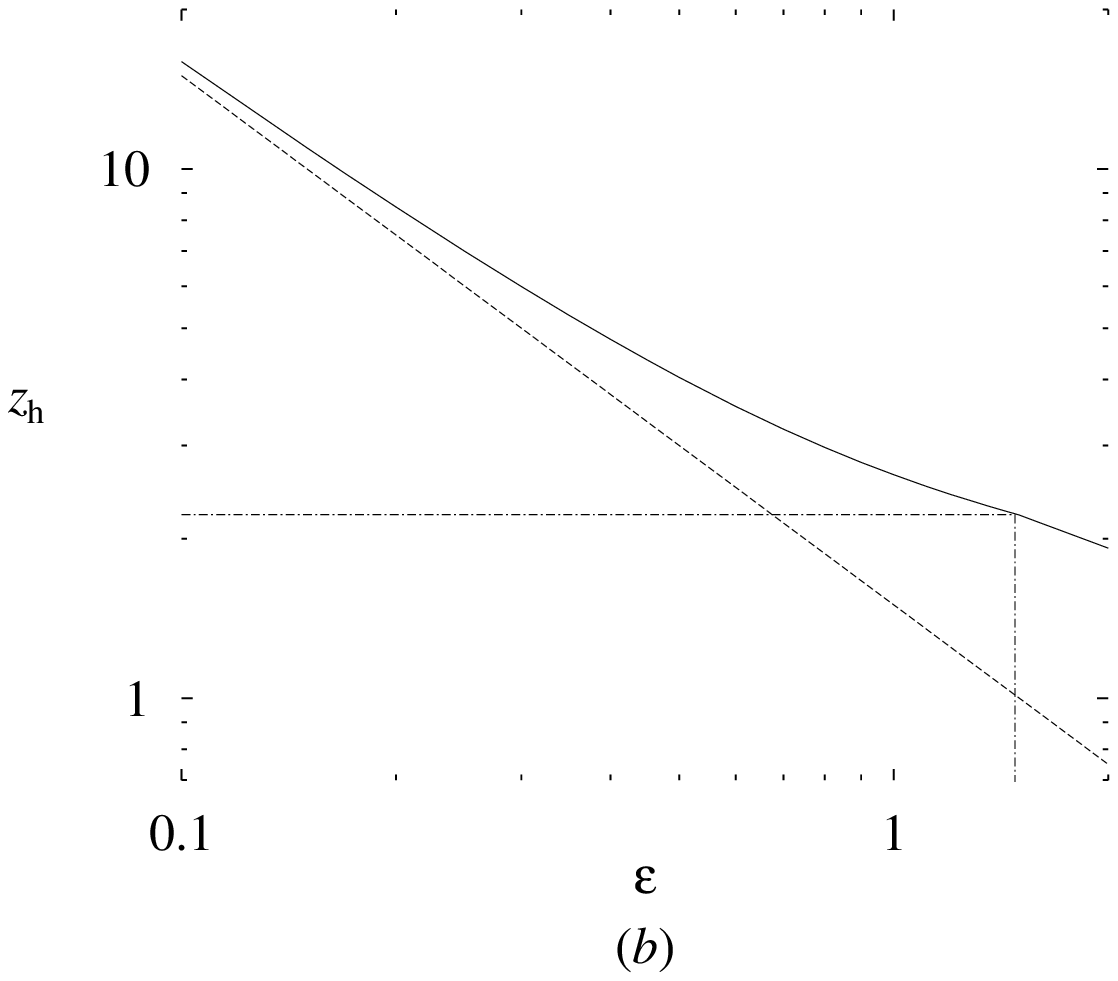,width=6.8cm}
  \end{center}
  \caption{({\it a\/}) The evolution of $X|_{z_{\rm h}}$ as a function of
           $\epsilon$. ({\it b\/}) Log-log plot of the proper distance to the
           horizon as a function of $\epsilon$ (solid line) compared with the
           first order prediction of $z_{\rm h} = 3 / 2 \epsilon$ (dashed
           line). The dash-dotted line indicates the phase transition at
           $\epsilon = 3/2, X|_{z_{\rm h}} \approx 2.221$.}
  \label{fig:phtrans}
\end{figure}

To determine whether there is indeed a phase transition in the behaviour
of the solutions, we examine the evolution of the value of the Higgs field at
the horizon, $X|_{z_{\rm h}}$, as a function of $\epsilon$. For a wall
solution, this value represents the maximum of the function $X$ (see for
instance figure~\ref{fig:soln}{\it a\/}), and a value $X|_{z_{\rm h}} = X_{\rm
F} = 0$ corresponds to the the false vacuum de Sitter solution. We expect
therefore that $X|_{z_{\rm h}}$ will drop from $1$ at $\epsilon=0$ to $0$ for
some $\epsilon$ in the range $[\epsilon_{\rm max}, {4\over3}\epsilon_{\rm
max}]$. According to our previous discussion, it is the solutions in an
intermediate range of $\epsilon$ which interest us most. We find
(figure~\ref{fig:phtrans}{\it a\/}) that the scalar field undergoes a phase
transition at the value $\epsilon = \epsilon_{\rm max} = 3/2$, in perfect
agreement with the prediction~(\ref{epsmax}), at which point wall solutions
cease to exist, and only the de Sitter configuration remains.

Figure~\ref{fig:phtrans}{\it b\/} shows the evolution of the proper distance to
the horizon as a function of $\epsilon$; for small $\epsilon$ this approaches
the first order prediction $z_{\rm h} \simeq 3 / 2\epsilon$ (dashed line), but
higher order corrections rapidly spoil the agreement. The proper distance to
the horizon at the phase transition can be predicted by the condition $\cos
(kz_{\rm h}) = 0$, which~--- with $k^2 = \epsilon_{\rm max} / 3 = 1/2$~---
implies $z_{\rm h} = \pi/\sqrt{2} \approx 2.221$.

Thus we see that the numerical work confirms the general analytic 
derivations given earlier in the section, and indicates that at 
$\epsilon_{\rm max}=3/2$, the domain wall solution disappears entirely.

\subsection{The Sine-Gordon Potential}
\label{sec:sG}

Consider now the periodic sine-Gordon potential, $ V(X) = \frac12 (1 + \cos
X) = \cos^{2} X/2$.  As before to zeroth order in $\epsilon$, one gets 
the usual sine-Gordon soliton
\be
\label{sg1}
X_0 = 4 \arctan \left(\e^{{z\over2}}\right) - \pi,
\ee
as shown in figure~\ref{fig:flatwall}.

Making use of (\ref{aone}, \ref{xone}), we obtain the gravitational
back reaction to order O$(\epsilon)$,
\bml
\label{sg2}
\bea
  A &=& 1 -4 \epsilon \; \ln \cosh {z\over2}\\
  X &=& 4 \arctan \left(\e^{{z\over2}}\right) - \pi 
-6\epsilon\, z \; \sech {z\over2}
\eea
\eml
and $k=2\epsilon+{\rm O}\left(\epsilon^2\right)$, which fixes the horizon
distance to $z_{\rm h} \simeq 1/2\epsilon$.
Note the agreement with equation (3.19) of Widrow's paper \cite{larryw}, who
also considered a sine-Gordon domain wall, although he did not compute
the correction to the scalar field.

Again, we must turn to numerical methods to find solutions for higher values of
$\epsilon$. The results we find are qualitatively very similar to those
obtained for $\lambda \Phi^4$. In particular we observe again the phase
transition predicted in the previous section for a general potential. This
time, however, the analytic results predict $\epsilon_{\rm max} = 3/16$ 
and $z_{\rm h} = \pi/2k
= 2\pi$; again, this is in excellent agreement with the numerical results.

\section{Domain walls with a cosmological constant}

In this section we consider the previous theories in a universe with a non-zero
cosmological constant $\Lambda$, which would correspond to gravitating domain
walls in an inflating universe. The effect of this constant can be readily
taken into account by modifying equations~(\ref{gweqs}) as follows:
\bml\bea
  X'' + 3 {A'\over A} X' &=& {1\over2} {\partial V\over \partial X} 
    \label{gweqbL} \\
  {A''\over A} &=& -{\epsilon\over3} \left [ 2 X^{\prime 2} + V(X) \right ] -
    \frac13 \Lambda \label{gweqaL} \\
  \left ( {A'\over A} \right )^2 &=& {k^2\over A^2} + {\epsilon\over3}
    \left [ X^{\prime 2} - V(X) \right ] - \frac13 \Lambda. \label{gweqcL}
\eea\eml
There are two qualitatively distinct cases: if $\Lambda > 0$, the wall is
embedded in a de Sitter background, whereas if $\Lambda < 0$ it is in an
anti-de Sitter background.\footnote{Strictly speaking, for an anti-de Sitter
background, in order to have the reflection symmetry around $z=0$ we need to
have $k^2<0$, which for a real metric would require $b(t) = \cos kt$. This in
turn requires the $\{x,y\}$ sections to be hyperbolic (see for example
\cite{CGS}); however, since this does not affect the 
equations of motion for $A(z)$, we will not discuss it further, and instead
refer the reader to \cite{CS}\ (and references therein)
for a detailed review of anti-de Sitter domain
walls.} The latter is of particular interest because the effect of the
cosmological constant should counteract the (effective) cosmological
constant created by the wall's back reaction.

First let us review how the analytic arguments of the previous section are
affected by a cosmological constant. Note that a false vacuum solution $X =
X_{\rm F}$ will now have an effective cosmological constant $\Lambda_{\rm eff}=
\Lambda + \epsilon$, hence the metric will be of the form (\ref{sitter}) with
$k^{2} = \Lambda_{\rm eff}/3$. Therefore, the previous arguments go through
essentially unchanged, but with $\Lambda_{\rm eff}$ instead of $\epsilon$.
Therefore
\be
\epsilon_{\rm max} = {3\over 8} |V''(X_{\rm F})| - \Lambda
\ee
Obviously, the range of the instability is increased for negative $\Lambda$,
and decreased for positive $\Lambda$. If $\Lambda > {3\over 8} |V''(X_{\rm
F})|$, then the false vacuum solution is `stable' (and indeed the only one).

\begin{figure}[htbp]
  \begin{center}
    \epsfig{file=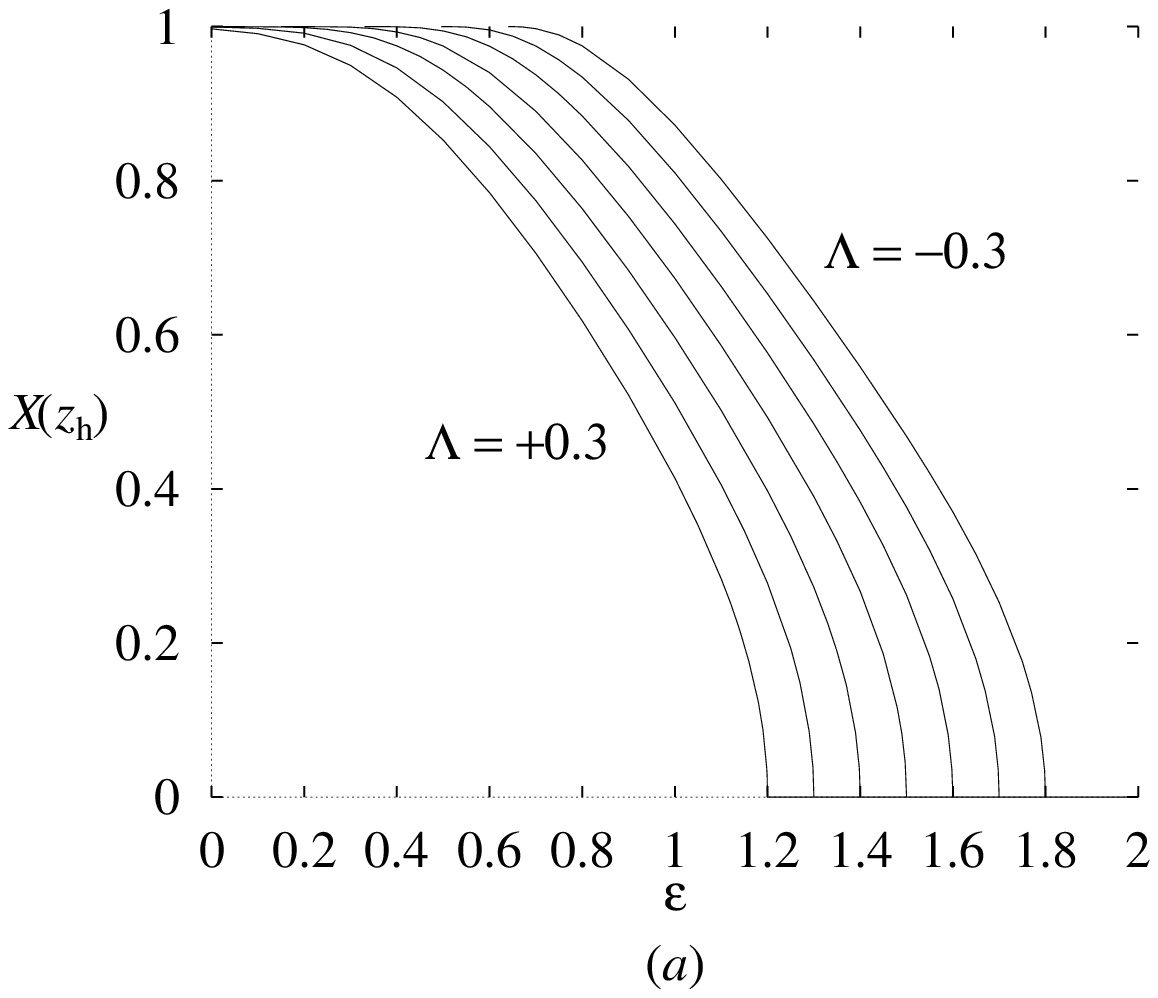,width=6.8cm} \qquad
    \epsfig{file=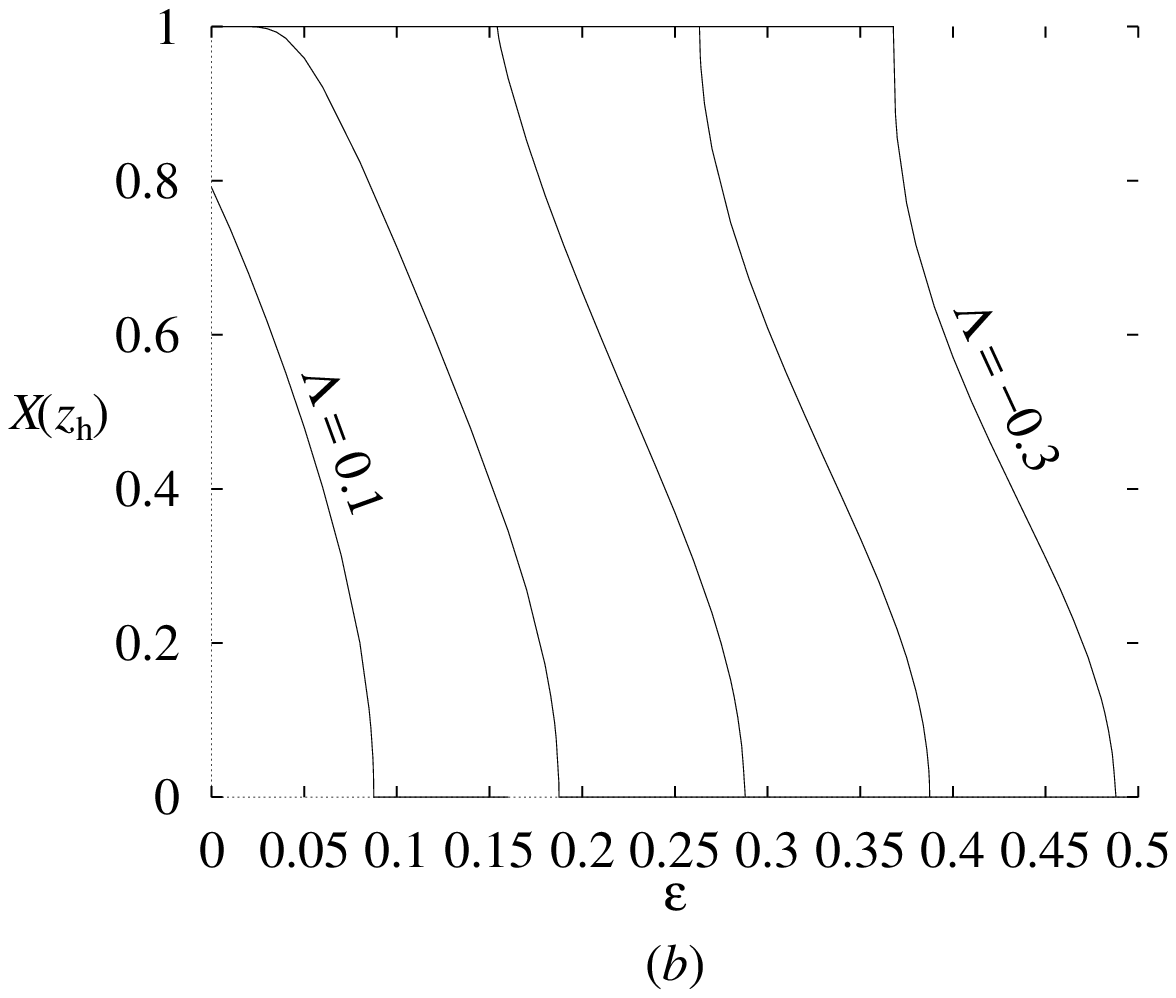,width=6.8cm}
  \end{center}
  \caption{Evolution of $X|_{z_{\rm h}}$ in function of $\epsilon$ and
           $\Lambda$ (from right to left, $\Lambda = -0.3, -0.2, \ldots
           0.2, 0.3$). ({\it a\/}) shows the $\lambda \Phi^4$ case, and ({\it
           b\/}) shows the sine-Gordon case. In ({\it b\/}) we have actually
           divided $X|_{z_{\rm h}}$ by $\pi$ to help the comparison with case
           ({\it a\/}).}
  \label{fig:hig}
\end{figure}

\begin{figure}[htbp]
  \begin{center}
    \epsfig{file=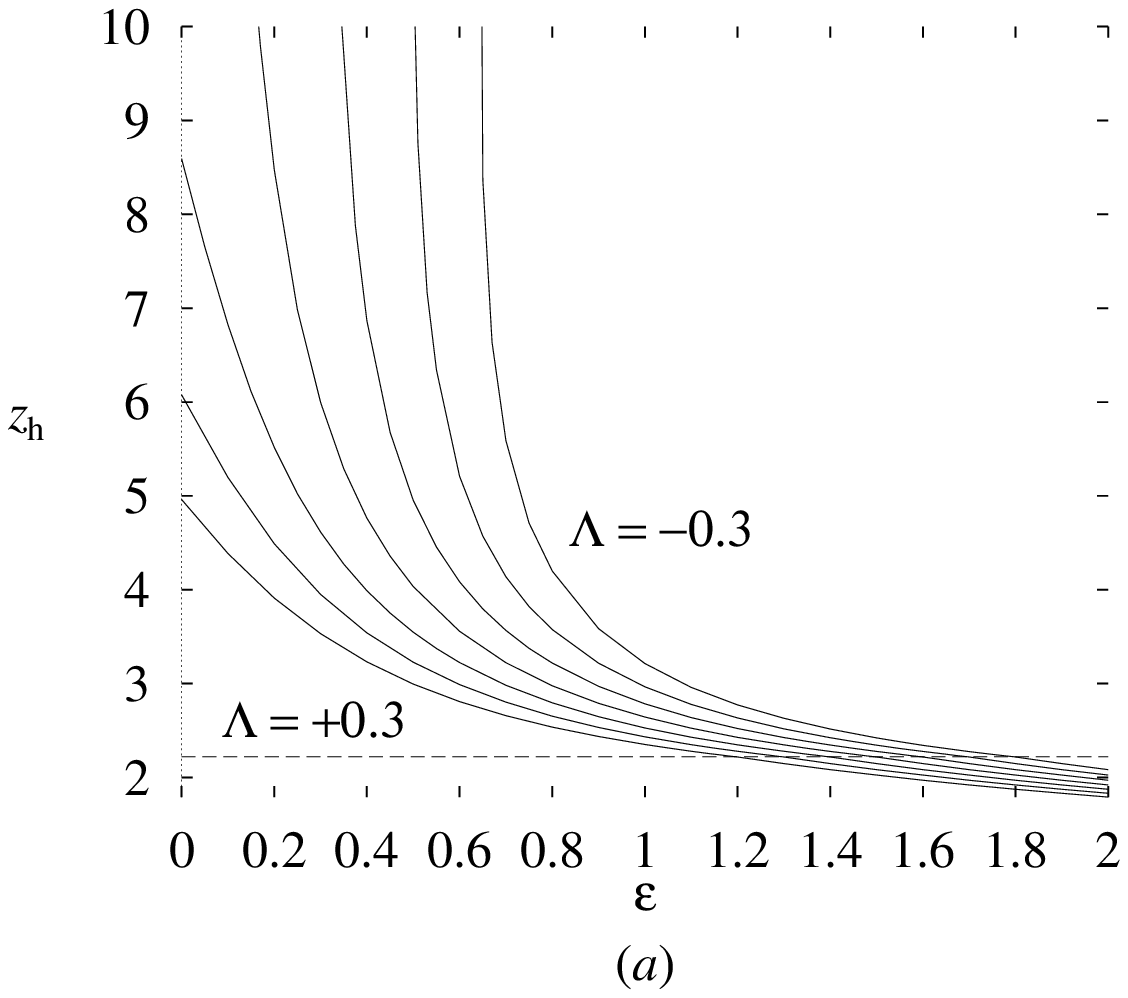,width=6.8cm} \qquad
    \epsfig{file=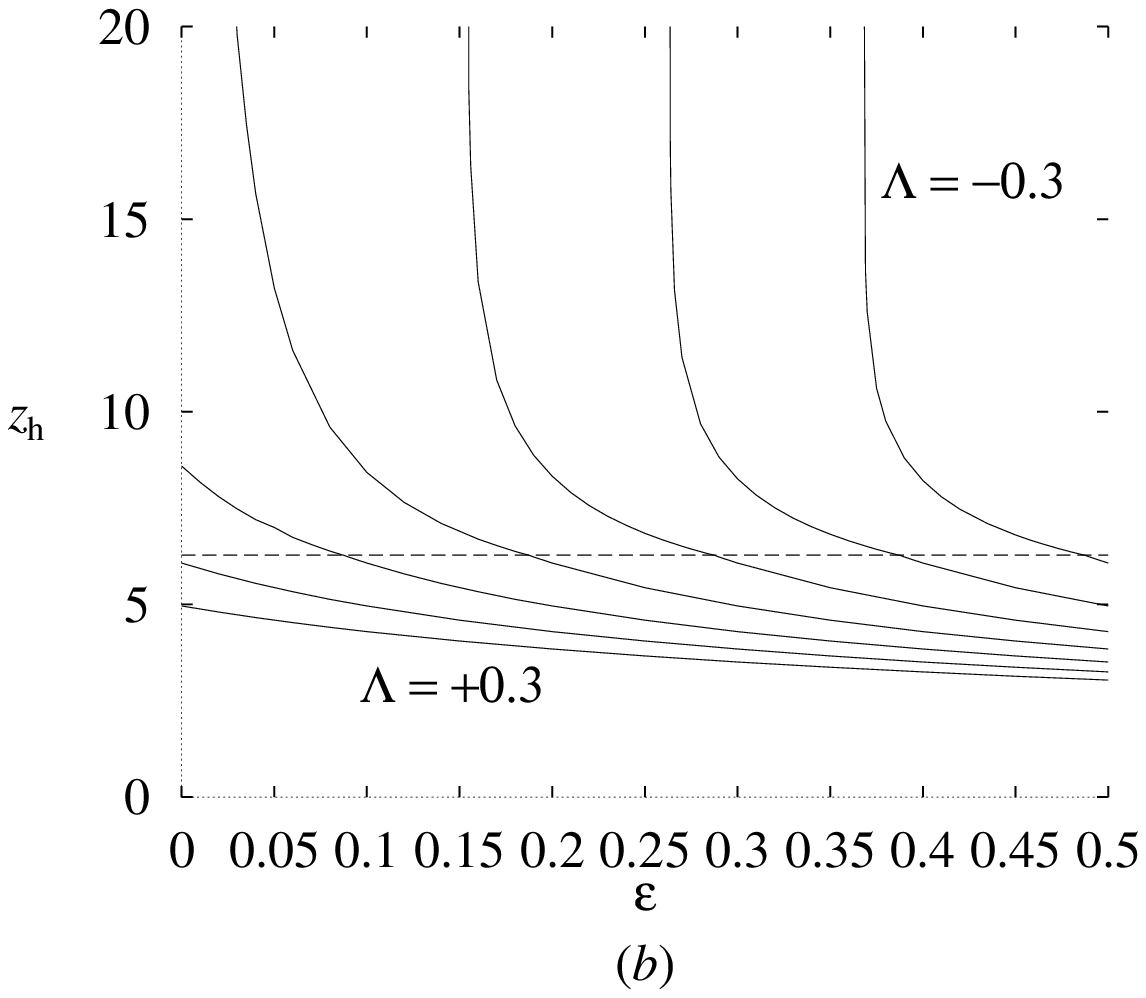,width=6.8cm}
  \end{center}
  \caption{Distance from the wall to the horizon, as a function of $\epsilon$
           and for the same values of $\Lambda$ as in figure~\ref{fig:hig}.
           ({\it a\/}) was obtained for $\lambda \Phi^4$ and ({\it b\/})
           for sine-Gordon. Again, the broken lines show the values of
           $z_\mscr{h}$ at the phase transition.}
  \label{fig:hor}
\end{figure}

This is illustrated in figures~\ref{fig:hig} and~\ref{fig:hor}, which show the
evolution of $X|_{z_{\rm h}}$ and $z_{\rm h}$ with $\epsilon$ for both $\lambda
\Phi^4$ and the sine-Gordon models as well as for several values of the
cosmological constant, $\Lambda = -0.3, -0.2, \ldots, 0.3$. In partiular, note
that for sine-Gordon (figure~\ref{fig:hig}{\it b\/}) the formula above tells us
that for $\Lambda > 3/16 = 0.1875$ the only solution is $X \equiv X_{\rm F}$;
this is why we do not see the curves for $\Lambda = 0.2$ and $0.3$.
Note as well that the value of $z_{\rm h}$ at which the phase transition occurs
($\pi / \sqrt{2}$ for $\lambda \Phi^4$, and $2\pi$ for sine-Gordon) remains
unaltered by the inclusion of the cosmological constant, as expected from the
discussion above. In fact, all the de Sitter solutions remain identical if
$\epsilon$ and $\Lambda$ are allowed to vary but $\Lambda_{\rm eff}$ remains
constant. (This is obviously not the case for the wall solutions, as $\epsilon$
then multiplies terms containing the Higgs field.)

\begin{figure}[htbp]
  \begin{center}
    \epsfig{file=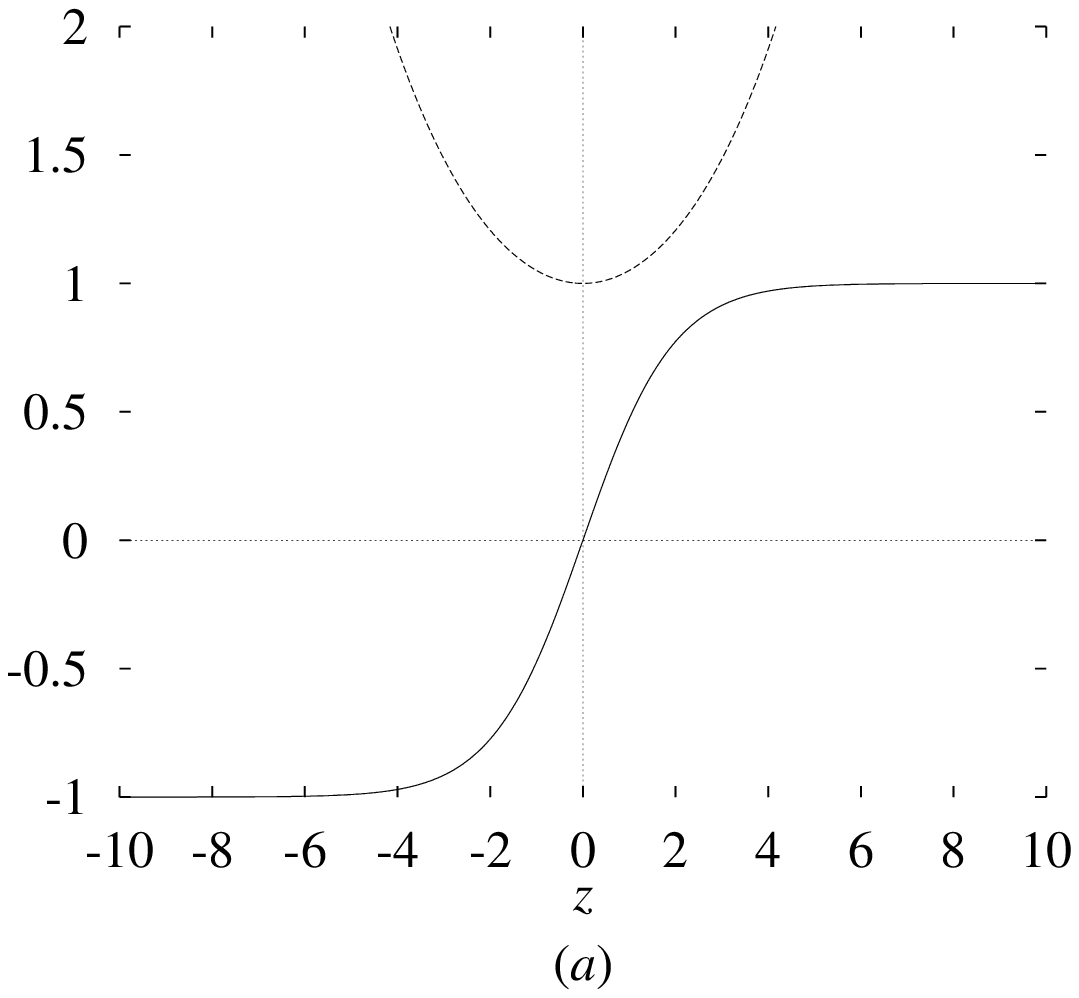,width=6.8cm} \qquad
    \epsfig{file=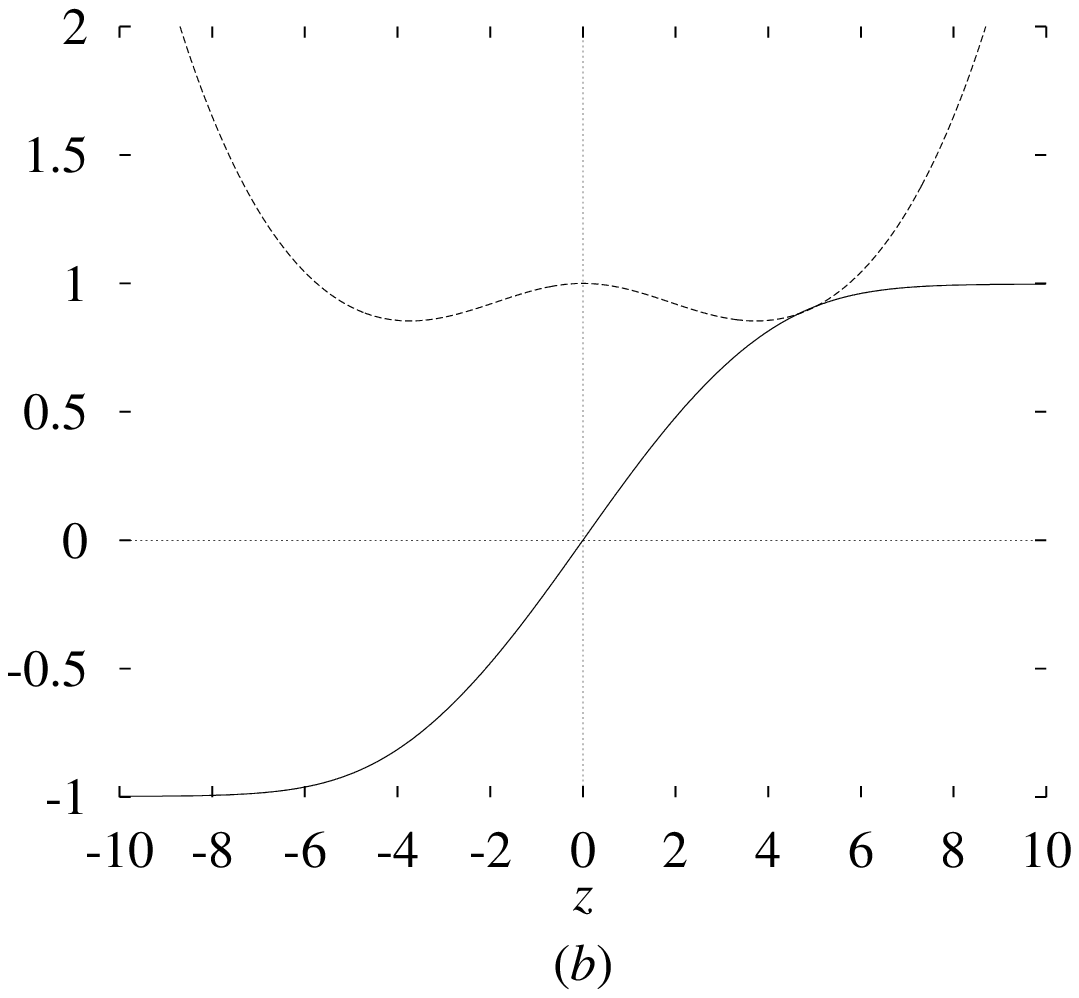,width=6.8cm} \\
    \epsfig{file=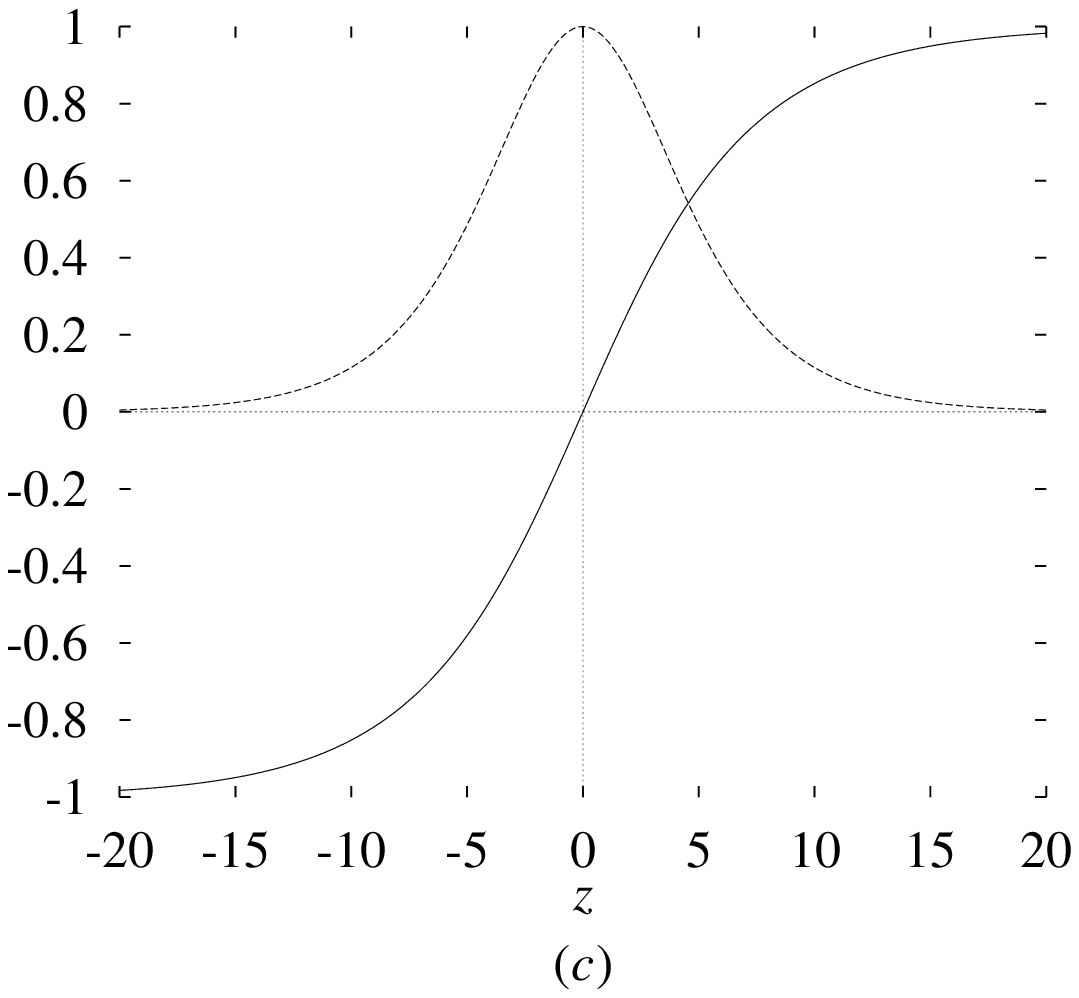,width=6.8cm} \qquad
    \epsfig{file=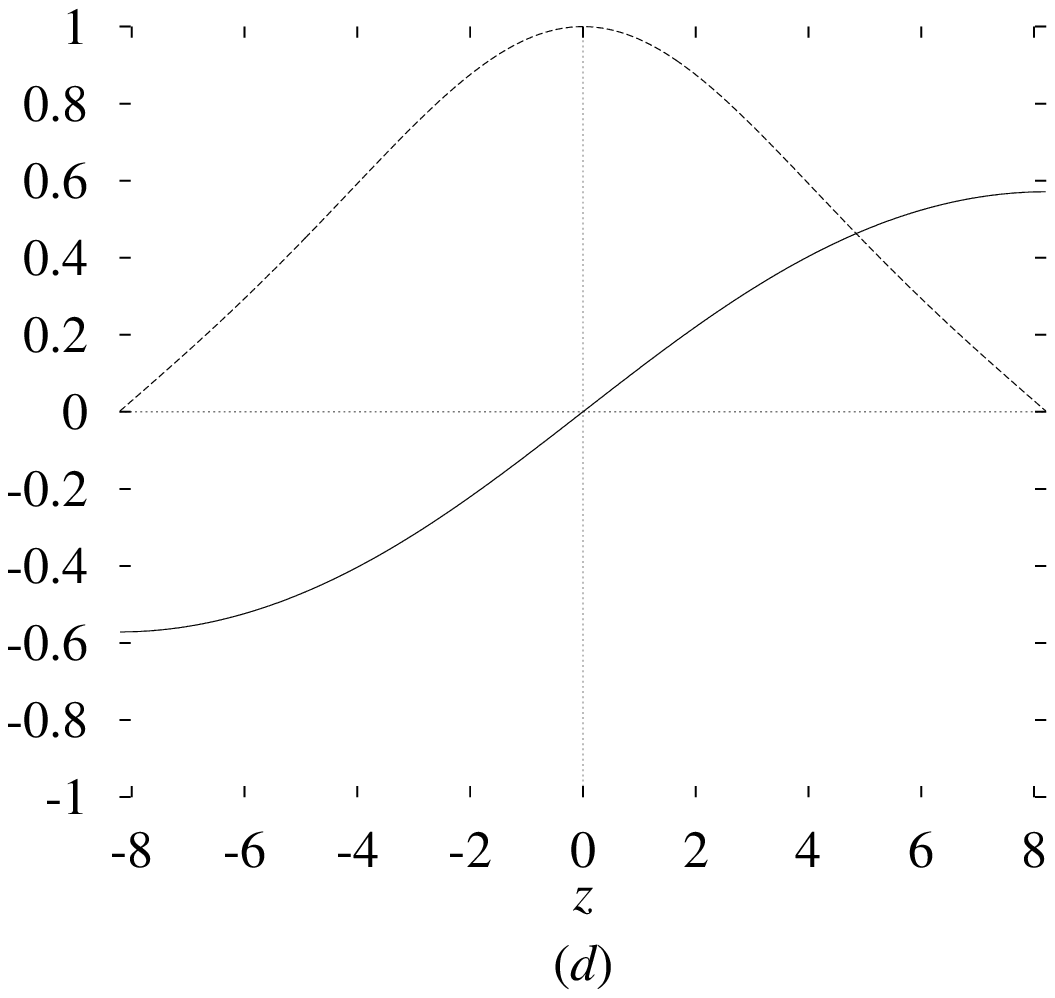,width=6.8cm}
  \end{center}
  \caption{Solutions $X(z)$ (solid lines) and $A(z)$ to the sine-Gordon
           equations for $\Lambda = -0.3$ and $\epsilon = $ 0 ({\it a\/}); 0.2
           ({\it b\/}); 0.367\ldots ({\it c\/}) and 0.4 ({\it d\/}).}
  \label{fig:seq}
\end{figure}

Now let us turn to the solution in the anti-de Sitter case, $\Lambda < 0$. We
now find \emph{three} qualitatively distinct solutions. For very small
$\epsilon$, the wall's self-gravitation cannot compete with the anti-de Sitter
expansion and $A'/A$ is strictly positive; in fact, it is easy to check that
the solution plotted on figure~\ref{fig:seq}{\it a\/} is $A(z) = \cosh(\sqrt{
|\Lambda|/3}\, z)$. As one increases $\epsilon$, the potential is observed to
decrease close to the wall's core, whereas the Higgs profile is slightly
smoothed~(figure \ref{fig:seq}{\it b\/}). In fact, this is the beginning of a
complete change in the metric function $A(z)$: as the wall's gravitational
interaction is switched on, $A$ assumes the shape of a ``double well,'' with a
local maximum at the imposed boundary value $A(0) = 1$ and two local minima
symmetrically situated at $A(\pm z_{\rm m})$ for some $z_{\rm m}$. As
$\epsilon$ increases, this double well becomes deeper, whereas $z_{\rm m}$
moves away from the wall. Notice that so far the function $A(z)$ is strictly
positive, and therefore none of these solutions exhibit an event horizon.
Eventually, however, for some critical value, $\epsilon_{\rm c}$, of
$\epsilon$, the two minima of $A(z)$ vanish as $z_{\rm m} \to \infty$
(figure~\ref{fig:seq}{\it c\/}). This would appear to be a thick wall version
of the type II extreme domain wall spacetime of Cvetic and Griffies \cite{CG},
and is therefore presumably supersymmetrizable. For $\epsilon > \epsilon_{\rm
c}$, the metric becomes negative at a finite distance $z_{\rm h}$, thus
giving rise to the wall's horizon.  The Goetz solution \cite{Goetz}\ lies in
this range.

\begin{figure}[htbp]
  \begin{center}
    \epsfig{file=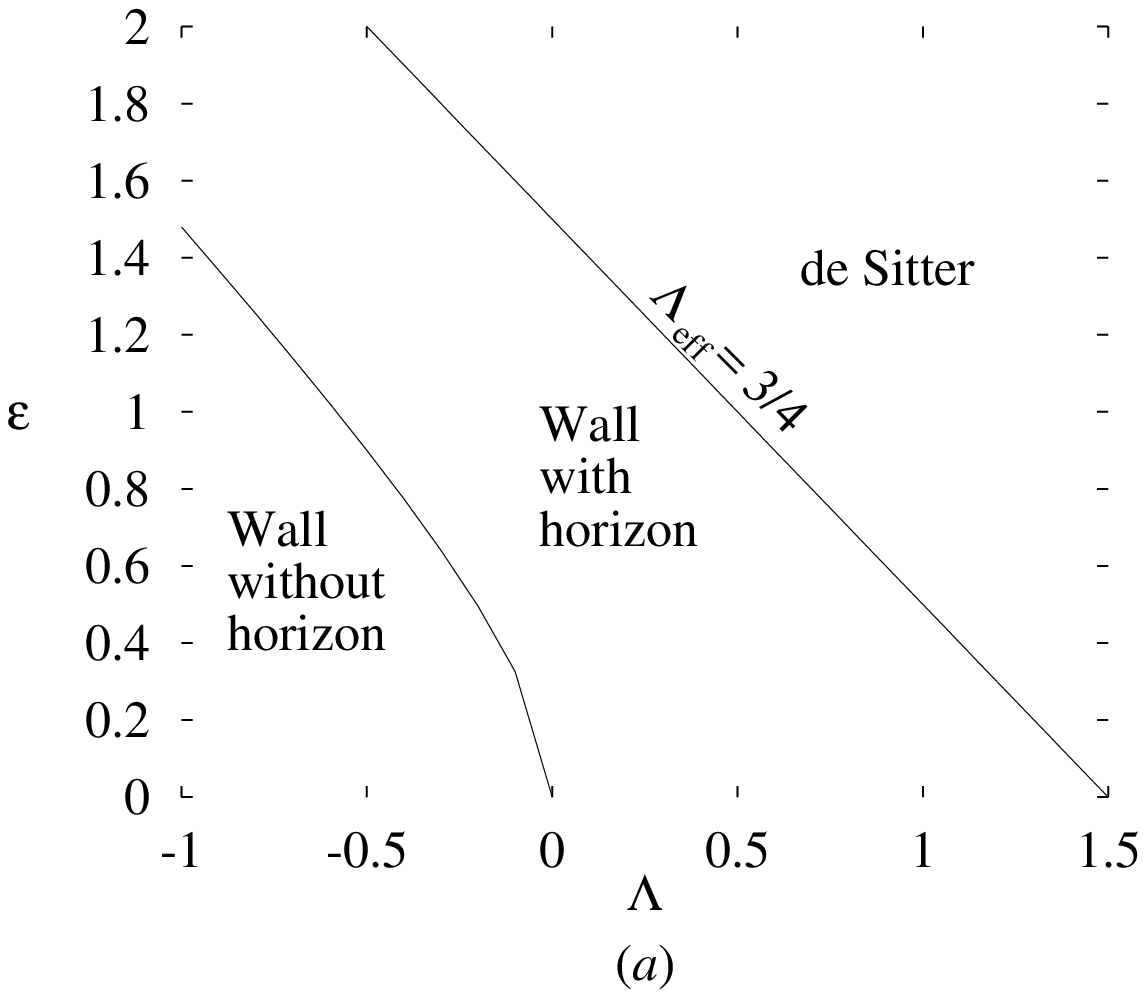,width=6.8cm} \qquad
    \epsfig{file=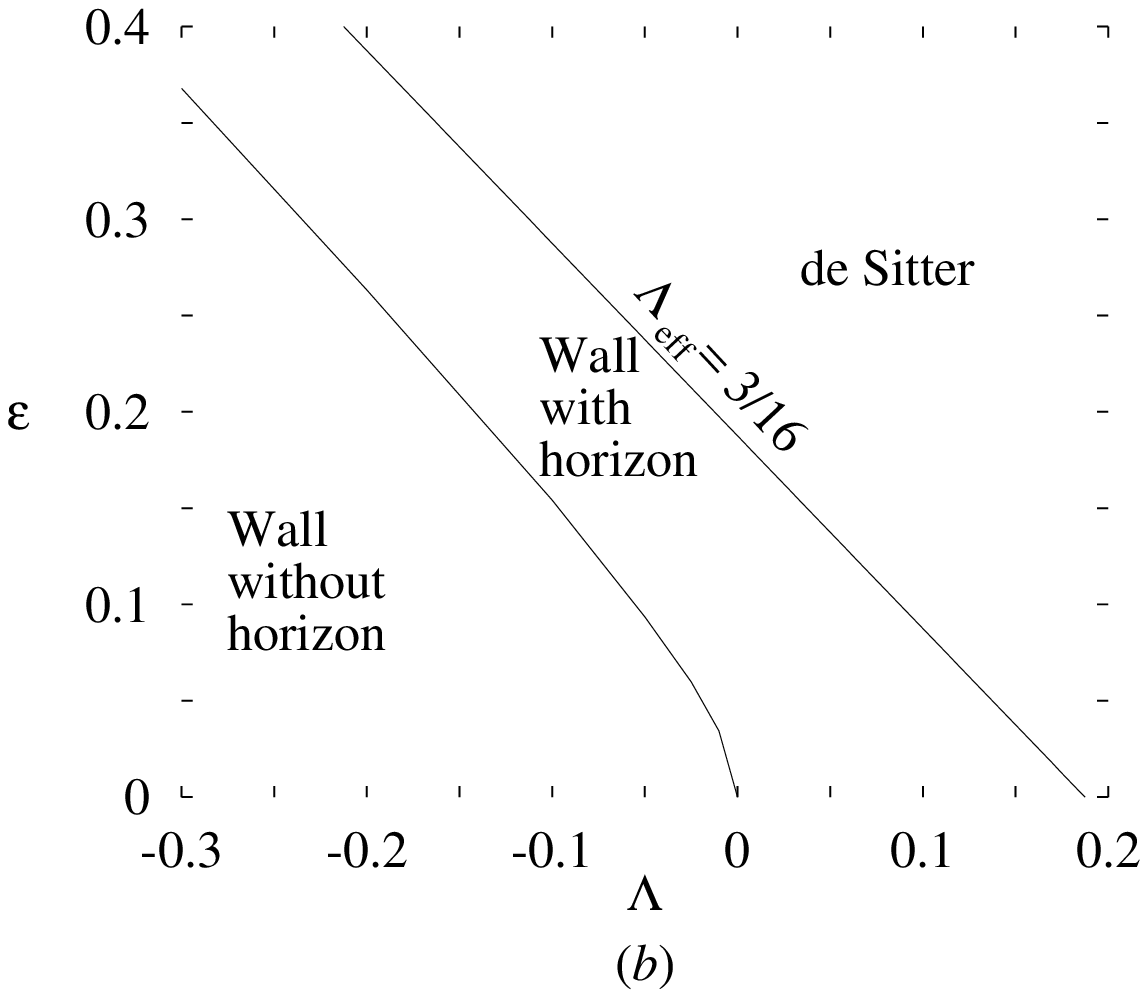,width=6.8cm}
  \end{center}
  \caption{Parameter space $(\epsilon, \Lambda)$ and the types of solutions
           found.}
  \label{fig:par}
\end{figure}

Figure~\ref{fig:par} shows the parameter space $(\Lambda, \epsilon)$, and the
different kinds of solution that we find. It is interesting to note that the
two lines separating the three phases seem to run parallel to each other in
both cases, indicating that a phenomenon similar to the triple point observed
in the phase diagram of water never occurs. This is to be expected, since as
long as the wall does not have an event horizon it is constrained to take its
asymptotic value at infinity. Of course, this topological constraint does not
imply that the lines are parallel, merely that they cannot meet in the physical
range $\epsilon > 0$; figure~\ref{fig:par} then shows that the range of the
parameter $\epsilon$ over which the value of the Higgs field at the horizon is
allowed to drop from 1 to 0 is fixed.

\section{Discussion}

To summarize, we have investigated the spacetimes of thick, gravitating domain
walls in detail, using both analytical arguments and numerical integration.
Both methods demonstrate the existence of a phase transition in the nature of
the `wall' solution, from being wall-like to a pure false vacuum de Sitter
spacetime. We find that for walls much thinner than their cosmological horizon,
the Vilenkin thin wall solution is a good description of the spacetime.  For
thicker walls the spacetime differs more markedly until finally there is only
the de Sitter solution.  The transition occurs abruptly, when the gravitational
strength of the wall is order of magnitude unity.  For walls less strongly
gravitating than this critical value, the spacetime has the appearance of a
gravitating domain wall, possibly lying within a background de Sitter or
anti-de Sitter universe.  For very strongly gravitating domain walls however,
the de Sitter universe is the only solution with the symmetries we have
imposed, the cosmological constant being provided by the false vacuum 
energy of the wall. Of
course this does not prove that the de Sitter solution is the {\it only}
solution to the coupled Einstein-scalar field equations; by transforming to
a global coordinate system for de Sitter, one can find a complete set of
solutions for the tachyonic wave equation, hence we can find an instability of
the required parity which is given in the notation of section~\ref{sgwall}\
by
\be
\xi = \sin (kz) 
[v(x,y,z,t)]^{-(\nu+5)/2} 
~_2{\rm F}_1 \left [ {\nu+1\over2},{\nu+5\over2},\nu +{7\over2} ; v\right]
\ee
where $\nu$ is given by (\ref{nudef}), and $v$ by
\be
v(x,y,z,t) = \left [ 1 + \cos^2kz \left ( \sinh kt + {\half} k^2 (x^2+y^2)
e^{kt} \right )^2 \right ] ^{-1}
\ee
and F is a hypergeometric function. However, since this instability 
depends on the $\{x,y\}$ coordinates, it does not correspond to decay to
a solution of the type considered in this paper,
and presumably corresponds to decay to a time dependent spherical wall
of the type found numerically by Sakai et.\ al.\ \cite{SSTM} which are
of relevance to topological inflation \cite{Linde,Vil2}. These
solutions are not solitons in the sense of having a fixed profile in 
time, and therefore are outside the consideration of this paper\footnote{
We would like to thank Alex Vilenkin for discussions on this point.}.

This phase transition behaviour of the scalar is reminiscent of the flux
expulsion of a vortex core by the horizon of an extremal black hole
\cite{BEG,CCES}. There, it was the fact that in the extremal limit the black
hole horizon $\cal H$ decoupled from the exterior spacetime that allowed a
partially analytic analysis of the vortex equations on the $S^2$ surface that
was the event horizon. The findings in that case, \cite{BEG}, parallel our
results here very closely. There was always a solution corresponding to the
fields taking their false vacuum values on $\cal H$, but for sufficiently thin
vortices (relative to the horizon radius) there was an additional solution
corresponding to a vortex anti-vortex pair at opposite poles (in the case of a
string threading the black hole). In that case however, the false vacuum
solution for thin strings could be shown not to extend to a full solution in
the exterior spacetime. Here, we always have a false vacuum de Sitter solution,
which is unstable to wall formation, as well as the defect solution for low
gravitational coupling. It is easy to see the common feature in these two
problems~--- the compact nature of the spatial section upon which the defect
must live. Defects on compact spaces have been analysed; for example, Avis and
Isham~\cite{AI} explored some years ago the $\lambda \Phi^4$ solutions on a
circle. They found exact solutions for the scalar field in terms of elliptic
functions, and no solution other than false vacuum if the radius of the circle
was too small. However, the crucial difference of our work to~\cite{AI}\ 
and~\cite{BEG} is that we are not looking at defects on a fixed background,
but looking for self-gravitating wall solutions without specifying their
topology \emph{ab initio}.

\begin{figure}[htbp]
  \begin{center}
    \epsfig{file=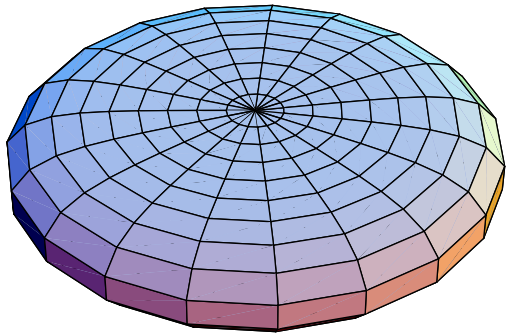,width=6.8cm}
  \end{center}
  \caption{The spatial section of the weakly gravitating sine-Gordon
           domain wall. The surface shown is the $t^* = z^*=0$ surface
           for $\epsilon = 1/30$.}
  \label{fig:discus}
\end{figure}

The topology of a black hole event horizon is obviously compact, however, it
turns out that in fact the topology of a domain wall spacetime is also
compact~\cite{GWG}. To see that not only de Sitter spacetime but also the
domain wall spacetime is topologically $S^3\times$\real, consider the
coordinate transformation (\ref{starred}). If we define a fifth coordinate,
$w^*$, by
\be
  \label{wstar}
  w^* = \int_0^z{\sqrt{1 - {A^{\prime2}\over k^2}} \; dz}
\ee
then the wall metric becomes a slice of a five dimensional flat metric, and we
can view our spacetime as a four-dimensional hypersurface embedded in
five-dimensional flat spacetime in an analogous fashion to the de Sitter
hyperboloid. From (\ref{starmet}) the equation for this hypersurface is
\be
  t^{*2} - {\bf x}^{*2} = - {A^2(w^*)\over k^2}.
  \label{surface}
\ee
For example, the de Sitter solution is $A = \cos kz$, therefore $w^* = {1\over
k} \sin kz$ from (\ref{wstar}), and (\ref{surface}) reduces to $t^{*2} - {\bf
x}^{*2} - w^{*2} = -1/k^2$, the de Sitter hyperboloid. For small $\epsilon$
on the other hand, the sine-Gordon wall from (\ref{sg2}) gives
\be
  w^* = \int_0^z \sech{z\over2} dz = 4 \arctan \e^{z\over2} - \pi
  =X_0(z), \label{wsg}
\ee
hence the hypersurface is given by a hyperboloid which has been deformed by
squashing in the $w^*$ direction,
\be
  \label{oonesurf}
  t^{*2}-{\bf x}^{*2} = -{1\over 4\epsilon^2} \left (
  1 + 4\epsilon \log \cos {w^*\over2}\right )^2.
\ee
The spatial section is depicted in figure~\ref{fig:discus} which shows the
$t^*=z^*=0$ slice to O($\epsilon$) for $\epsilon = 1/30$.

The spatial geometry of the domain wall is therefore topologically $S^3$, and
is similar to a discus, although the upper and lower surfaces are flat almost
to the edge of the discus. This corresponds exactly to our intuitive idea of
the spacetime exterior to the wall being flat, with a highly localised region
of curvature generated by the wall itself located at the rim of the discus. It
fleshes out the thin wall description of the spacetime, smoothing over the
distributional singularity of the thin wall hypersurface.  The two length
scales of the domain wall spacetime correspond to the two radii of the discus,
and as $\epsilon$ increases, the discus radius shrinks, with its height
remaining much the same, until the geometry is almost spherical, at which point
the radius becomes too small to support a defect solution and we make the
transition to a false vacuum de Sitter hyperboloid, with an exactly spherical
spatial geometry.
What is interesting in comparison with the Avis and Isham scenario is that it
is the self gravity of the domain wall which produces a compactification of
spacetime at its own characteristic scale, which ultimately becomes too small
to support the defect itself.

\begin{figure}
  \centerline{\epsfig{file=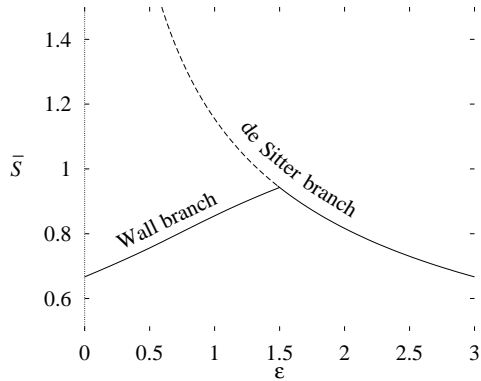,width=6.8cm}}
  \caption{Bifurcation diagram for the gravitating domain wall. We plot here
           the normalized action $\bar{S}$ as a function of $\epsilon$.
           Solid lines indicate the stable solutions found numerically, and the
           dashed line represents the unstable de Sitter solution.}
  \label{fig:bifurc}
\end{figure}

Finally, in analogy with \cite{AI}, we can make a plot of
the normalized action for the domain wall solution versus the de Sitter
solution
\be
  S = \int {\cal L}_{\rm G} + {\cal L}_{\rm M} = \eta^2 \int V(X) \, \e^{2kt}
      A^3 d^4x = N \int_0^{z_{\rm h}} A^3 V(X) dz,
\ee
where $N$ is a normalization factor, and there are no boundary terms from the
gravitational part of the action.  For the false vacuum de Sitter solution
$\bar{S} = S/N = 2/3k = 2/\sqrt{3\epsilon}$, and for the weakly gravitating
$\lambda\Phi^4$ model, ${\bar S} = 2/3$.  Figure~\ref{fig:bifurc} shows a plot
of the action of the $\lambda \Phi^4$ wall against the false vacuum de Sitter
solution, which indicates clearly the instability of the latter solution to
wall formation for $\epsilon < 3/2$. (This can be compared to the 
instanton action plot obtained in \cite{BV2}.)

\section*{Acknowledgements}
 
We would like to thank Alex Vilenkin and Richard Ward for helpful
discussions.
F.B. is supported by an ORS award and a Durham University award.
C.C. is supported by EPSRC and by the `R\'egion Centre.'
R.G. is supported by the Royal Society.
 
\def\apj#1 #2 #3.{{\it Astrophys.\ J.\ \bf#1} #2 (#3).}
\def\cmp#1 #2 #3.{{\it Commun.\ Math.\ Phys.\ \bf#1} #2 (#3).}
\def\comnpp#1 #2 #3.{{\it Comm.\ Nucl.\ Part.\ Phys.\  \bf#1} #2 (#3).}
\def\cqg#1 #2 #3.{{\it Class.\ Quant.\ Grav.\ \bf#1} #2 (#3).}
\def\jmp#1 #2 #3.{{\it J.\ Math.\ Phys.\ \bf#1} #2 (#3).}
\def\mpla#1 #2 #3.{{\it Mod.\ Phys.\ Lett.\ \rm A\bf#1} #2 (#3).}
\def\ncim#1 #2 #3.{{\it Nuovo Cim.\ \bf#1\/} #2 (#3).}
\def\npb#1 #2 #3.{{\it Nucl.\ Phys.\ \rm B\bf#1} #2 (#3).}
\def\phrep#1 #2 #3.{{\it Phys.\ Rep.\ \bf#1\/} #2 (#3).}
\def\plb#1 #2 #3.{{\it Phys.\ Lett.\ \bf#1\/}B #2 (#3).}
\def\pr#1 #2 #3.{{\it Phys.\ Rev.\ \bf#1} #2 (#3).}
\def\prd#1 #2 #3.{{\it Phys.\ Rev.\ \rm D\bf#1} #2 (#3).}
\def\prl#1 #2 #3.{{\it Phys.\ Rev.\ Lett.\ \bf#1} #2 (#3).}
\def\prs#1 #2 #3.{{\it Proc.\ Roy.\ Soc.\ Lond.\ A.\ \bf#1} #2 (#3).}


\begin{references}
\bibitem{VS}      A.Vilenkin and E.P.S.Shellard, {\it Cosmic strings and
                  other topological defects\/} (Cambridge: Cambridge University
                  Press 1994)
\bibitem{ABR}     A.Albrecht, R.A.Battye and J.Robinson, \prl 79 4736 1997. 
                  [astro-ph/9707129]
\bibitem{PST}     U.L.Pen, U.Seljak and N.Turok, \prl 79 1611 1997.
                  [astro-ph/9704165]
\bibitem{FMG}     P.Ferreira, J.Magueijo and K.M.Gorski, \apj 503 L1 1998.
                  [astro-ph/9803256]\hfill\break
                  J.Pando, D.Valls-Gabaud and L.Fang, \prl 81 4568
                  1998. [astro-ph/9810165] \hfill\break
                  D.Novikov, H.Feldman and S.Shandarin, [astro-ph/9809238].
\bibitem{Israel}  W.Israel, \ncim 44B 1 1966.
\bibitem{R}       R.Gregory, \prd 54 4955 1996. [gr-qc/9606002]
\bibitem{Vil}     A.Vilenkin, \plb 133 177 1983.
\bibitem{IS}      J.Ipser and P.Sikivie, \prd 30 712 1984.
\bibitem{GWG}     G.W.Gibbons, \npb 394 3 1993.
\bibitem{RD}      D.Garfinkle and R.Gregory, \prd 41 1889 1990. 
\bibitem{larryw}  L.M.Widrow, \prd 39 3571 1989. 
\bibitem{HSF}     C.T.Hill, D.N.Schramm and J.N.Fry, \comnpp 19 25 1989.
\bibitem{Goetz}   G.Goetz, \jmp 31 2683 1990. 
\bibitem{Mukh}    M.Mukherjee, \cqg 10 131 1993. 
\bibitem{Tom}     K.Tomita \plb 244 183 1990.
\bibitem{TI}      S.Tadaki and H.Ishihara, \prd 41 3047 1990.
\bibitem{Linde}   A.Linde, \plb 237 208 1994. [astro-ph/9402031]
                  \hfill\break
                  A.Linde and D.Linde, \prd 50 2456 1994. [hep-ph/9402115]
\bibitem{Vil2}    A.Vilenkin, \prl 72 3137 1994. [hep-th/9402085]
\bibitem{BV1}     R.Basu and A.Vilenkin, \prd 50 7150 1994. [gr-qc/9402040]
\bibitem{BBGH}    D.Boyanovsky, D.E.Brahm, A.Gonzalez-Ruiz, R.Holman and
                  F.I.Takakura, \prd 52 5516 1995. [hep-ph/9501380]
\bibitem{BV2}	  R.Basu and A.Vilenkin, \prd 46 2345 1992.
\bibitem{BGV}	  R.Basu, A.H.Guth and A.Vilenkin, \prd 44 340 1991.
\bibitem{BEG}     F.Bonjour, R.Emparan and R.Gregory, [gr-qc/9810061]
\bibitem{NumRec}  W.H.Press, S.A.Teukolsky, W.T.Vetterling and
                  B.P.Flannery, {\it Numerical recipes in C: The art of
                  scientific computing}, Second edition (Cambridge: Cambridge
                  University Press 1992)
\bibitem{CGS}     M.Cvetic, S.Griffies and H.H.Soleng, \prd 48 2613 1993.
                  [gr-qc/9306005]
\bibitem{CS}      M.Cvetic and H.H.Soleng, \phrep 282 159
                  1997. [hep-th/9604090]
\bibitem{CG}      M.Cvetic and S.Griffies, \plb 285 27 1992. [hep-th/9204031]
\bibitem{SSTM}	  N.Sakai, H.Shinkai, T.Tachizawa and K.Maeda, \prd 53
655 1996. Erratum \prd 54 2981 1996. [gr-qc/9506068]
\bibitem{CCES}    A.Chamblin, J.M.A.Ashbourn-Chamblin, R.Emparan and
                  A.Sornborger, \prl 80 4378 1998. [gr-qc/9706032] \prd 58
                  124014 1998. [gr-qc/9706004] \hfill\break
                  F.Bonjour and R.Gregory, \prl 81 5034 1998.
                  [hep-th/9809029]
\bibitem{AI}      S.J.Avis and C.J.Isham, \prs 363 581 1978.
\end{references}
\end{document}